\documentclass[iop,a4paper,fleqn,usenatbib]{mnras}%

\usepackage{newtxtext,newtxmath}  
\usepackage[T1]{fontenc} 
\usepackage{ae,aecompl}   

\usepackage{graphicx}
\usepackage{amssymb}
\usepackage{footnote}

\usepackage{setspace}
\usepackage[squaren, Gray, cdot]{SIunits}

\usepackage{makecell} 

\title[Manifold learning and galaxy properties]{{\sc Horizon-AGN} virtual observatory -- 2.\\
Template-free estimates of galaxy properties from colours.}

\author[I.~Davidzon et al. ]{
\parbox[t]{\textwidth}{I.~Davidzon$^{1,2}$\thanks{E-mail: iary.davidzon@nbi.ku.dk},  
C.~Laigle$^{3}$, 
P.\,L.~Capak$^{1,2}$, 
O.~Ilbert$^{4}$,  
D.\,C.~Masters$^{5}$, 
S.~Hemmati$^{5}$, 
N.~Apostolakos$^{6}$, 
J.~Coupon$^{6}$,
S.~de\,la\,Torre$^{4}$,
J.~Devriendt$^{3}$, 
Y.~Dubois$^{7}$,
D.~Kashino$^{8}$, 
S.~Paltani$^{6}$,
C.~Pichon$^{7,9}$
}
\vspace*{6pt} \\ 
$^{1}$ IPAC, Mail Code 314-6, California Institute of Technology, 1200 East California Boulevard, Pasadena, CA 91125, USA \\
$^{2}$ Cosmic Dawn centre (DAWN), Niels Bohr Institute, University of Copenhagen, Juliane Maries vej 30, DK-2100 Copenhagen, Denmark\\
$^{3}$ Sub-department of Astrophysics, University of 
Oxford, Keble Road, Oxford OX1 3RH, UK\\
$^{4}$ Aix Marseille Univ, CNRS, CNES, LAM, Marseille, France \\
$^{5}$ Jet Propulsion Laboratory, California Institute of Technology, Pasadena, CA 91109, USA \\
$^{6}$ Department of Astronomy, University of Geneva, ch. d'Ecogia 16, 1290, Versoix, Switzerland  \\
$^{7}$ Sorbonne Universit{\'e}s, CNRS, UMR 7095,  Institut d\'Astrophysique de Paris, 98 bis bd Arago, 75014 Paris, France \\
$^{8}$ Department of Physics, ETH Z{\" u}rich, Wolfgang-Pauli-strasse 27, CH-8093 Z{\" u}rich, Switzerland \\
$^{9}$ Korea Institute for Advanced Study (KIAS), 85 Hoegiro, Dongdaemun-gu, Seoul, 02455, Republic of Korea\\
}
\date{Accepted --. Received --; in original form --} 
\pubyear{2019}  
\begin{document}
\label{firstpage}
\pagerange{\pageref{firstpage}--\pageref{lastpage}}
\maketitle

\begin{abstract}
 Using the \textsc{Horizon-AGN} hydrodynamical  simulation and self-organising maps (SOMs), we show how to compress the complex, high-dimensional data structure of a simulation into a 2-d grid, which greatly facilitates the analysis of how galaxy observables are connected to intrinsic properties.  We first verify the tight correlation between the observed 0.3$\!-\!5\mu$m broad-band colours of \textsc{Horizon-AGN} galaxies and their high-resolution spectra. The correlation is found to extend to physical properties such as redshift, stellar mass, and star formation rate (SFR). 
 This direct mapping from colour to physical parameter space still works after including photometric uncertainties that mimic the COSMOS survey.   
 We then label the SOM grid with a simulated calibration sample to estimate redshift and SFR for COSMOS-like galaxies up to $z\sim3$. 
 In comparison to  state-of-the-art techniques based on synthetic templates, our method is comparable in performance but less biased at estimating redshifts, and significantly better at predicting SFRs.  In particular our ``data-driven'' approach,  in contrast to model libraries, intrinsically allows for the complexity of galaxy formation  and can handle sample biases.  We advocate that  observations to calibrate this method should be one of the goals of next-generation galaxy surveys.
\end{abstract}

\begin{keywords} 
galaxies: evolution, fundamental parameters -- methods: statistical,data analysis 
\end{keywords}

\section{Introduction}
\label{sec:introduction}

One of the most successful techniques to understand galaxy formation is measuring galaxy properties in large-area surveys and comparing the results with cosmological-scale simulations based on theoretical models of galaxy formation.  Typically the comparison is done in physical parameter space, so secure estimates of  redshift, luminosity ($L$), stellar mass ($M$), star formation rate (SFR), must be obtained from observational data.  These estimates usually come from the analysis of the spectral energy distribution (SED) or the high-resolution spectrum of galaxies, relying on the correlation between specific wavelengths and physical properties: for example H$_{\alpha}$ emission and star formation \citep{kennicutt98} or $\sim\!2\mu$m light and stellar mass \citep{madau98}.  In the past two decades data from multi-wavelength photometry and spectroscopic surveys has become abundant, and fitting galaxy templates to observed SED and spectra is now the standard method to perform this analysis \citep[among the pioneering studies:][]{sawicki&yee98,bell&dejong00,gavazzi02,PerezGonzalez2003,Fontana:2004p8883,gallazzi05}. Galaxy parameters are usually derived from the maximum-likelihood template \citep{bolzonella00} or from the full probability distribution function (PDF) of the template set \citep{benitez00BPZ}. 
For quantities like stellar mass or star formation history (SFH), templates are built from stellar population synthesis models \citep[e.g.,][]{bruzual&charlot03,maraston05,Conroy2009a}. 

The modelling process often introduces systematic effects that have been shown to severely bias $M$ and SFR estimates in some cases \citep[e.g.][]{mitchell13,mobasher15,laigle19}.   For instance, the synthetic templates are not guaranteed to have fully realistic features: e.g., their SFH is often an analytically function \citep[as the $\tau$- and inverted-$\tau$ models, see][]{maraston10} that does not include either multiple bursts or chemical enrichment.  Discrepancy between synthetic templates and real galaxies is also due to the assumptions about their stellar population initial mass function (IMF) and dust attenuation of their stellar light  \citep[e.g.,][]{davidzon13}. Moreover, templates not always take into account nebular emission lines, which may contaminate the observed broad band photometric colours.  Finally, some galaxy types may not even be included in the library  \citep[e.g.\ the old and dusty galaxies at $z\sim2\!-\!3$ discussed in][]{marchesini10}.

Beyond modelling problems, there are additional systematics introduced by the fitting procedure. The relative abundance of a given SED in the real universe is often not accounted in the synthetic library: most of the fitting codes assume that all templates are equally likely.   
Moreover the SED (or spectral) fitting algorithms may not treat the template set in an optimal way, as they often rely on computationally expensive brute-force approaches to explore the entire library \citep[see][]{speagle16} or they may introduce systematics when convolving templates with instrumental errors  \citep[see][]{cappellari17-pPXF}. 
These uncertainties all propagate into the commonly used statistical descriptors of a galaxy census, such as the galaxy stellar mass function \citep{ilbert13,grazian15} and the specific SFR evolution of star forming galaxies  \citep{santini17,davidzon18}. Ultimately, all of these uncertainties combine to result in a biased view of galaxy demographics, preventing a clear and straightforward comparison between observations and simulations. 

To make such a comparison more robust, significant effort has been devoted to improving template fitting techniques. To date, substantial progress has been achieved in each step of the computation, from the construction of galaxy models with more complex SFHs \citep{pacifici13} to improved radiative transfer modes including the interstellar medium \citep{dacunha08} and sophisticated Bayesian fitting techniques \citep[][]{leja17,chevallard16}.   

In parallel to this ongoing effort other authors have explored alternate paths, replacing standard template fitting with new techniques based on  machine learning (ML). Besides implicitly accounting for biases, a key advantage of ML techniques is indeed their speed, which enables analyses of extremely large data sets.  Most of the existing work aims at estimating redshifts (see \citealp{salvato19} and references therein), with the exception of a few publications where ML has been applied e.g.\ to recover SFR \citep{delliveneri19} and specific SFR  \citep{StensboSmidt17} of $z\sim0$ galaxies from the Sloan Digital Sky Survey (SDSS). A wider range of galaxy parameters (including stellar mass and metallicity) is estimated in \citet{simet19}. That study, complementary to ours,  use a \textit{supervised} neural network trained with a semi-analytic model simulation. 

In the present work we describe a novel technique  based on \textit{unsupervised} ML combined with analytic data modelling, to simultaneously provide  redshift and SFR estimates for  galaxies across a large redshift range ($0<z\lesssim3$, spanning about 12\,Gyr of universe's life). 
The ML algorithm adopted here is the self-organising map  \citep[SOM,][]{kohonen81}, which is an unsupervised manifold learning algorithm used to analyse high-dimensional data \citep[see also][]{kohonen01}. Initially popular in engineering research, it soon circulated to many other  fields including Astrophysics. 
Seminal work has used the SOM mainly to classify astronomical objects and their properties, including stellar populations  \citep{hernandez-pajares94}, star vs galaxy separation \citep{miller&coe96,maehoenen&hakala95},  and morphological types  \citep{molinari&smareglia98}. Since it does not require the manifold to have physical meaning, the SOM has also been used to classify astronomical publications \citep[][]{poincot98}. More recently the SOM has been applied to calibrating redshifts for weak lensing cosmology \citep{masters15}. Other recent studies using the SOM will be mentioned throughout this work. 

As other ML methods, the SOM starts with a training phase.  However, unlike supervised methods, the goal of the training is to create a compressed, lower-dimensional representation of the data rather than estimate an output. In this work we first perform the training on galaxy colours drawn from the \textsc{Horizon-AGN} simulation. 
We then label the SOM with galaxy properties not learned during the training phase.  These labels can be drawn from a data model based either on simulations or, as suggested later, \textit{bona fide} galaxies observed as reference for calibration.  Since the mapping from the data to the labels is explicit and analytic, control over selection functions, sample biases, and the effects of observational noise is retained (unlike supervised ML where these factors are part of the learning scheme). 

In this paper we focus on galaxy SFR estimates because they are fundamental to constraining galaxy evolution \citep{madau&dickinson14} along with stellar mass measurements. However, compared to the latter, the SFR estimates from template fitting are much more uncertain: previous work \citep[e.g.,][]{laigle19} shows that SFR is more sensitive  than stellar mass to SED-fitting assumptions. This sensitivity is inherent to relying on the ultraviolet (UV) continuum as a star formation indicator because it is highly attenuated by dust, geometry dependent, and also sensitive on the details of SFH.  Other techniques offer better performance by using either far-infrared (FIR)  data \citep[$\geqslant$24\,$\mu$m,][]{lefloch09,dacunha08} or spectroscopic follow-up \citep{kennicutt98,Kewley2004} which is expensive and impractical to obtain for every galaxy. Therefore we focus the present study on estimating SFR from the rest-frame UV to near-IR (NIR) photometry, leaving the details of other physical properties to future work.  

To test and develop our ML method we use a mock galaxy catalogue of $\sim\!8\times10^5$ objects extracted from the cosmological hydrodynamical simulation \textsc{Horizon-AGN} \citep{dubois14}; the mock catalogue was presented in \citet[][hereafter Paper I]{laigle19} as the first milestone of the \textsc{Horizon-AGN} virtual observatory project.  
One of the main goals of the project is to bridge the divide between empirical and theoretical studies by adding observational-like features to simulated galaxy samples. 
To this purpose, we produced  mock catalogues with  characteristics similar to the COSMOS survey \citep{Scoville:2007p13732} and \emph{Euclid} \citep[as predicted in][]{Laureijs:2011vu}. By applying our SOM estimator to the COSMOS-like version of the \textsc{Horizon-AGN} galaxies we aim to demonstrate its feasibility for real data sets. 
Both the simulated data set and the SOM are described in Sect.~\ref{sec:data&methods}.

Building on this result, in Sect.~\ref{sec:estimator} we introduce the SOM-based estimator of redshift and SFR, and apply it to the COSMOS-like mock catalogue. In Sect.~\ref{sec:real_calibr} we show that redshift and SFR must be known for only a  subset of galaxies in order to ``calibrate'' the SOM estimator and provide estimates without a synthetic template library.  We then discuss  possible ways to build such a calibration sample, inspired by previous work proposing a highly complete spectroscopic follow-up in large cosmological surveys    \citep{masters15,hemmati19}.
To show the improvement of template-free ML with respect to standard fitting techniques, Sect.~\ref{sec:real_calibr} ends with a comparison between SOM estimates vs the SFR derived in Paper I for the same galaxies by means of the code \textsc{LePhare} \citep{arnouts99,ilbert06}. 
We discuss the results and draw our conclusion in Sect.~\ref{sec:conclusion}.  

Throughout this work we use a flat $\Lambda$CDM cosmology with 
$H_{0}=70.4$\,km\,s$^{-1}$\,Mpc$^{-1}$,  
$\Omega_{m}=0.272$,  $\Omega_{\Lambda}=0.728$,  and $n_s=0.967$ \citep[][WMAP-7]{komatsu11}. 
All magnitudes are  in the AB \citep{Oke:1974p12716} system. The IMF is as in \citet{chabrier03}. 



\begin{table}
    \caption{Terminology used throughout this work (see also Sect.~\ref{subsec:intro_som}).}
    \label{tab:lexicon}
    \begin{tabular}{cl}
    \textit{Parameter space} & \makecell[l]{$M$-dimensional space where each dimension\\ is a galaxy feature (a colour in our case).}\\[8pt]
    \textit{Colour} ($\mathcal{C}$) & Observer's frame colour (e.g., $u-g$). \\[4pt]
    SOM & \makecell[l]{Self-organising map connecting the $M$-d\\ space to a lower dimensional space ($N$-d).} \\[8pt]
     \textit{Grid} & \makecell[l]{Low-dimensional space with rectangular\\ shape ($N\!=\!2$) used here to display the SOM.} \\[8pt]
     \textit{Cell} & \makecell[l]{Minimum element of the grid, 
    which can\\ contain one or more galaxies.}\\[8pt]
    \textit{Weight} ($\hat{w}$) & \makecell[l]{$M$-d vector for a given cell, connecting it \\to a point in the parameter space.}\\[8pt]
    \textit{Training} & \makecell[l]{Iterative process to define the weights\\ and distribute galaxies into cells.}\\[8pt]
    \textit{Mapping} & \makecell[l]{Projecting a new galaxy on the grid,\\linking it to its best-matching weight/cell.}\\[8pt]
    \textit{Calibration} & \makecell[l]{Assign a value (label) to each cell,\\ according to any galaxy property.}\\[8pt]
    $\langle \,\dots\, \rangle^\mathrm{cell}$ & \makecell[l]{Median of a generic quantity computed for\\ galaxies in the same cell.}
    \end{tabular}
\end{table}

\section{Data and Methods}
\label{sec:data&methods}

\subsection{The \textsc{Horizon-AGN} virtual observatory}
\label{subsec:intro_simu}

The present study relies on a mock galaxy catalogue built from the {\sc Horizon-AGN} hydrodynamical simulation\footnote{
\url{http://www.horizon-simulation.org/}}
\citep[][]{dubois14}. This catalogue, presented in \citet[][]{laigle19}, includes 789,354 galaxies extracted from a $1\times1$ deg$^2$ lightcone by running the \textsc{ AdaptaHOP} halo finder \citep{aubert04} on the stellar particles distribution. Each stellar particle (of mass $\sim 2\times 10^{6}\,M_{\odot}$) is linked to a synthetic simple stellar population  \citep[][hereafter BC03]{bruzual&charlot03}  assuming Chabrier's IMF \citep[][]{chabrier03} and interpolating between the metallicity values available in BC03.  The galaxy catalogue is  selected  in stellar mass  ($M_\mathrm{sim}>10^9\,M_\odot$) and redshift\footnote{
 cosmological redshifts ($z_\mathrm{sim}$) in our light-cone include galaxies' peculiar velocity.}
($0<z_\mathrm{sim}<4$).
The upper limit in redshift is imposed to work with galaxy colours consistently defined across the whole redshift range, i.e.\ avoiding non-detection in the $u$ band due to the Lyman limit shifting into the band. 

Our virtual observatory mimics the optical and near-infrared (NIR)  photometry of COSMOS2015 galaxies \citep{laigle16} in 10 broad bands ($u$, $B$, $V$, $r$, $i^{++}$, $z^+$, $Y$, $J$, $H$, $K_\mathrm{s}$)  and 14 medium-band filters \citep[from Subaru/SuprimeCam, see][]{taniguchi07}. It also includes the   \emph{Spitzer}/IRAC channels centred at 3.6 and 4.5\,$\mu$m (hereafter $[3.6]$ and $[4.5]$). In each filter we reproduce the signal-to-noise ratio ($S/N$) distribution as in the ``ultra-deep'' stripes of COSMOS2015. Reference  3$\sigma$ limits (in 3$\arcsec$ apertures) used in previous work are $K_\mathrm{s}<24.7$ and $i^+<26.2$ \citep[for a list of sensitivity depths in every filter, see Table 1 in][]{laigle16}. After introducing such uncertainties we perturb the original galaxy fluxes accordingly. Attenuation by dust and inter-galactic medium is also taken into account, whereas flux contamination by nebular emission is not implemented. In the following we refer to the attenuated fluxes without photometric errors as  \textit{intrinsic}, while \textit{perturbed} photometry is the one that takes into account galaxy $S/N$.  

Further details about the realisation of the \textsc{Horizon-AGN} mock galaxy catalogue can be found in Paper~I and Appendix A1 (see On-line Supplementary Materials). 
We also use a simpler (``phenomenological'') simulation to show that neither the Horizon-AGN limit in stellar mass nor the absence of nebular emission lines affect our main results (Appendix A2). 

\begin{figure*}
    \centering
    \includegraphics[width=1.1\columnwidth]{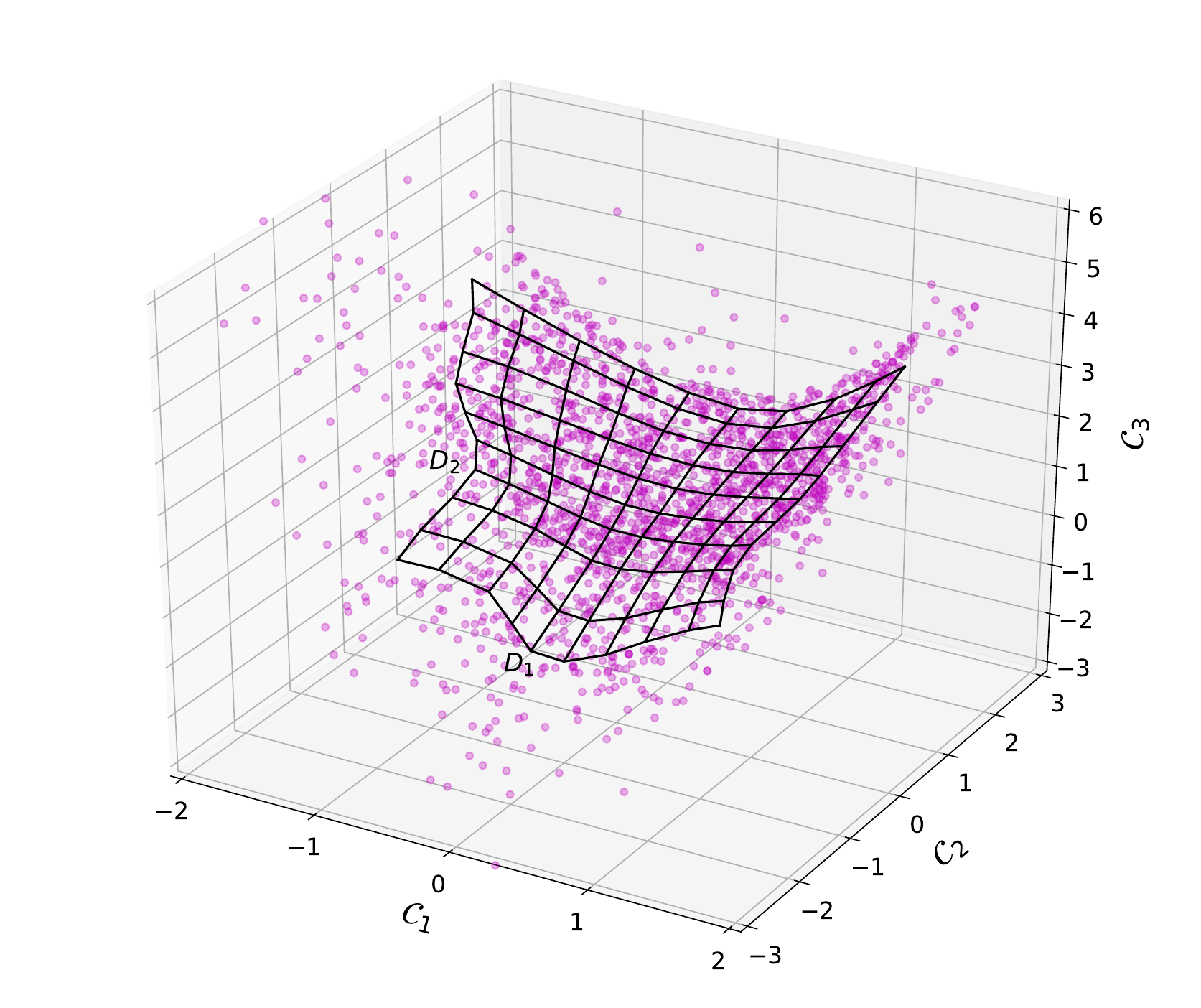}\\
    \includegraphics[width=0.55\columnwidth]{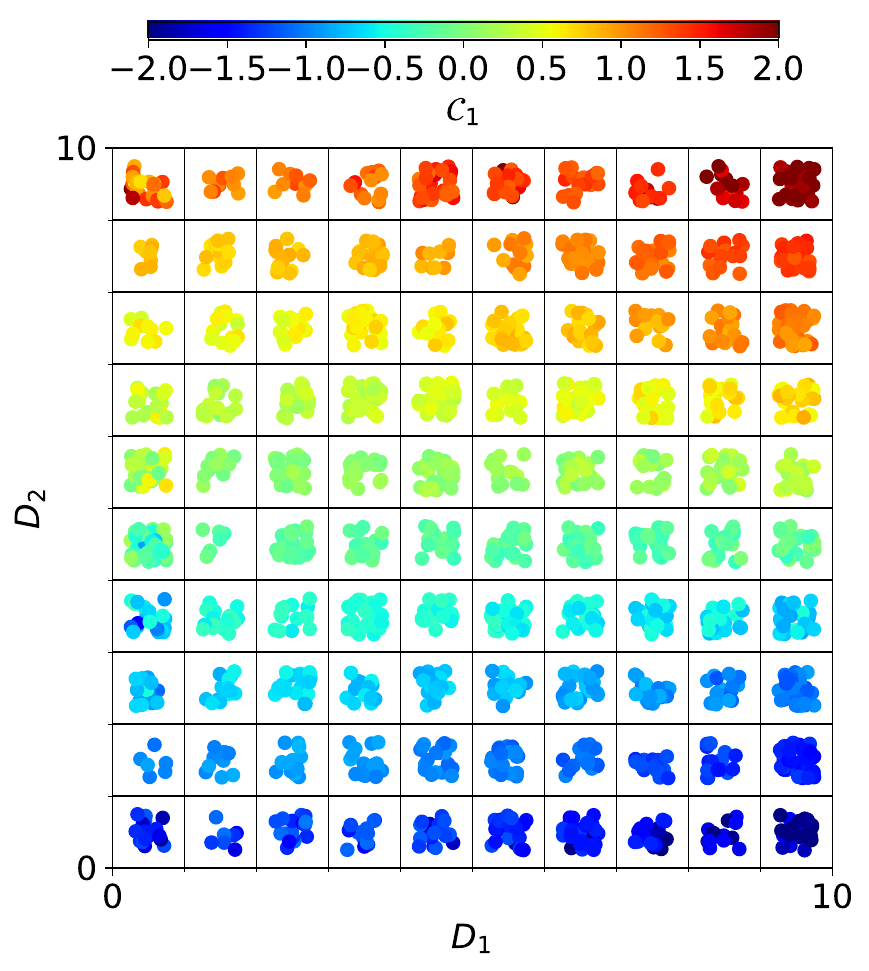}\hspace{10mm}
    \includegraphics[width=0.55\columnwidth]{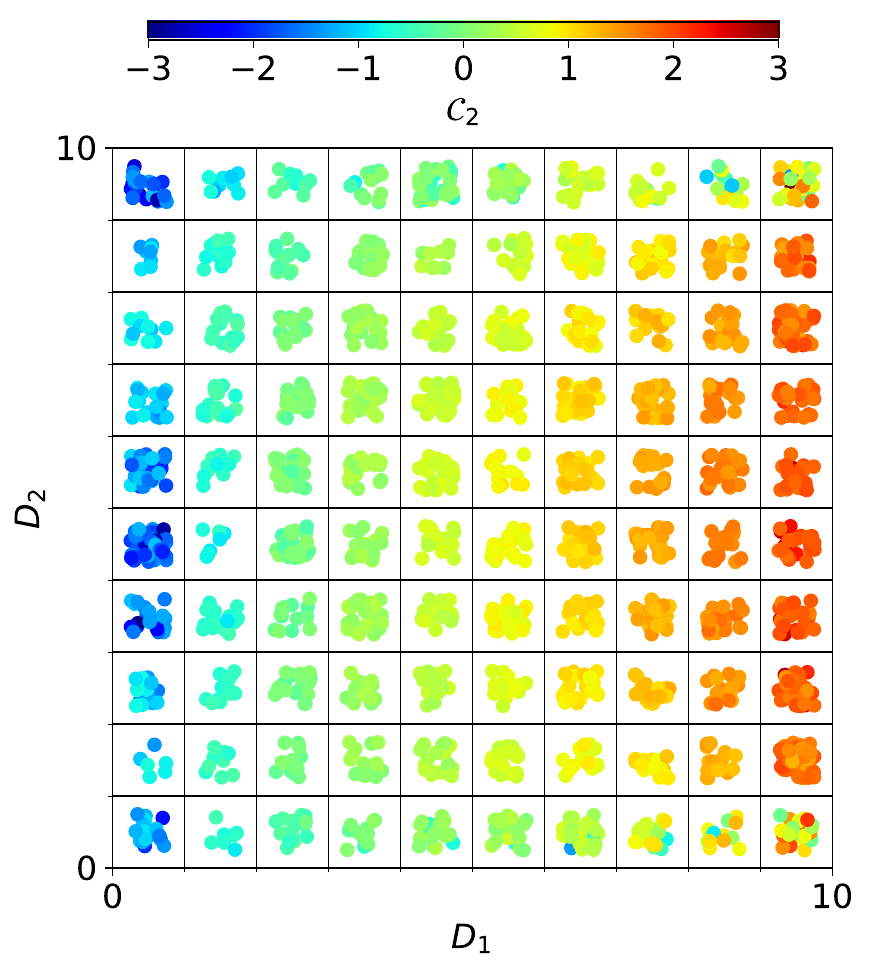}\hspace{10mm}
    \includegraphics[width=0.55\columnwidth]{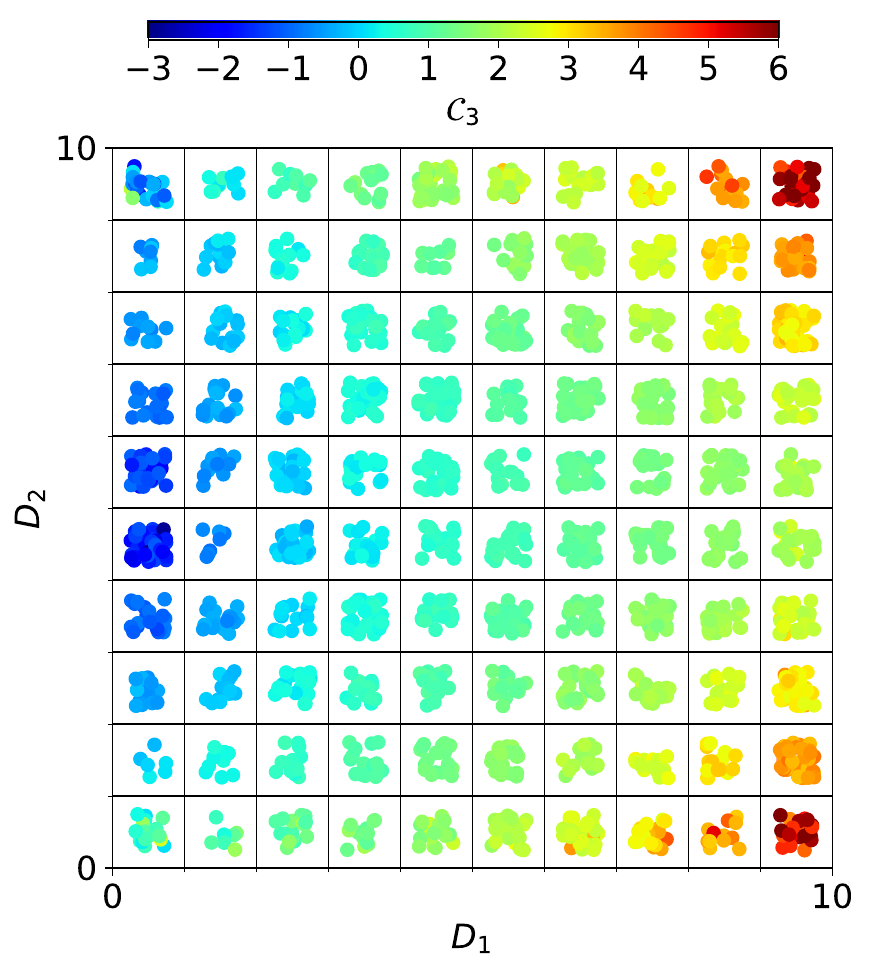}
    \caption{Pedagogical representation of the SOM using an artificial distribution of 2,000 objects in an unspecified 3-D space.  \textit{Upper panel:} After setting a $10\times10$ grid of ``cells'' the algorithm adapts the grid (black lines) to the manifold of the training  objects (magenta points). Here, as well as in the next figures, SOM axes are labelled with conventional names $D_1$ and $D_2$.  The grid becomes finer where the density of the training sample is higher, while sparse objects in the outskirt will be linked to the border of the grid. Features $\mathcal{C}_1, \mathcal{C}_2, \mathcal{C}_3$ can be regarded as three galaxy colours, but in principle they can be to any feature. \textit{Lower panels:} The SOM as it appears in the 2-d space, with a different layout in the three panels.  The 2,000 training objects (small dots) have been allocated into the $10\times10$ grid (their position within a cell has been scattered for illustrative purposes). In each panel objects are colour-coded according to one of their features ($\mathcal{C}_1, \mathcal{C}_2, \mathcal{C}_3$ from left to right). Similar objects are clustered in the same (or nearby) cell and a smooth transition is observed across the grid. Outliers with extreme characteristics (e.g., $\mathcal{C}_3>4$) are pushed to the grid corners.  
    In this example we use a simple distribution of data points for illustrative purposes. As a consequence the SOM grid is a ``simple'' 3-d surface.  In general the grid can assume more complex high-dimensional configurations. Thanks to this property the SOM and similar non-linear dimensionality reduction algorithms can accurately map the parameter space of real galaxies (which are a non-linear manifold).  This is also a key difference from PCA, which can only assumes a hyper-surface. 
    }
    \label{fig:som_didact}
\end{figure*} 

\subsection{The self-organizing maps}\label{subsec:intro_som}

In brief, a self-organising map (SOM) represents a high-dimensional data distribution into fewer dimensions (e.g., a 2-D space) through an \textit{unsupervised} neural network that preserves topology. In other words, objects that are multi-dimensional neighbours remain close to each other also in the 2-D space. 

Assume that the original (compact) space $\mathcal{M}$, with dimensions equal to $M$, has to be reduced into a space $\mathcal{N}$,   which we choose to be bi-dimensional ($N=2$)\footnote{Choosing $N=3$ or higher is possible but it would not offer the same advantage in terms of visualisation.}. To build the SOM we create a neural network where  each neuron is associated to a weight vector $\hat{w}$. Each element in these vectors  comes from the corresponding dimension of $\mathcal{M}$ (i.e., $\hat{w}$ has length $M$).   Neurons (and the vectors attached to them) are ordered in the $N$-dimensional configuration defined by the user, for instance a rectangular  lattice (see a pedagogical example in Fig.~\ref{fig:som_didact}). 

The SOM relies on a \textit{training sample} of objects drawn from $\mathcal{M}$. The neural network explores $\mathcal{M}$ by adapting neurons' weights to the training sample. Such a learning phase proceeds by iteration until the value of each weight gets as close as possible (according to a convergence  criterion) to the input data. 
The first task in the procedure is to find the \textit{best-matching unit} for any given data point ($\hat{x}$) of the training sample.  The best matching unit is the neuron whose weight $\hat{w}_b$ is the closest from $\hat{x}$. Then, the weight of each neuron (including the best-matching unit) is updated during an iterative process:  
\begin{equation}
    \hat{w}_i(t+1) = 
    \hat{w}_i(t) + \alpha(t)\,\phi(\hat{w}_i,\hat{w}_b,t)
    \left[ \hat{x} - \hat{w}_i(t) \right]. 
    \label{eq:w_update}
\end{equation}
Equation (\ref{eq:w_update}) is written for the $i$-th neuron, updated from step $t$ to $t+1$. The learning coefficient $\alpha$ is a monotonically decreasing function to ensure convergence, while $\phi$ is a  neighbourhood function that modulates the update depending on the distance between the $i$-th neuron and the best-matching unit:
    $\phi \propto \exp(-\vert\hat{w}_i-\hat{w}_b\vert)$. 
The procedure is repeated by scanning the other elements of the training sample.     
It is critical to use a training sample that is representative of the whole space, otherwise the grid of neurons/weights is adjusted to probe only a sub-set of $\mathcal{M}$. 

The learning process is unsupervised because it does not require the training sample to be labelled \textit{a priori}. Neurons autonomously organise their weights: 
hence the name ``self-organizing'' map. The resulting SOM is a mapping function that connects a point from $\mathcal{N}$ to $\mathcal{M}$ and vice versa. We stress out that the topology is preserved so that in the low-dimensional  configuration  (the 2-D lattice in our example) two adjacent weights are linked to nearby regions of $\mathcal{M}$.

In our case the high-dimensional space is the typical baseline of an extragalactic survey, i.e.\ each dimension is a colour measured from broad-band filters (for example $u-g$, $g-r$, etc.). The training sample is a set of galaxies large enough to span the colour distributions observed in the universe. We specifically explore the panchromatic space because there is a straightforward connection to the physics of galaxy evolution we aim at studying\footnote{
A colour is more informative  about galaxies' SFH than single broad-band magnitudes. 
The SOM can also analyse more complex high-dimensional spaces defined by a combination of miscellaneous parameters (galaxy colours, fluxes, morphology, etc.).}. We choose a rectangular lattice to order the neurons. Because of its appearance, and to maintain the same lexicon of previous work, hereafter  we will refer to the neurons as ``cells'' in a 2-D ``grid''  (Fig.~\ref{fig:som_didact})\footnote{
It should be noticed that in our SOM implementation input data will be normalised in each dimension, re-scaling the distribution to unit variance and centring the mean at zero.}. 
Each cell is defined by its weight vector, for example $\hat{w}_{ij}$ for the cell with coordinate $i,j$ in the grid. 
The weight connects its cell to a point in the panchromatic space, i.e.\ 
the vector components $(w_{ij,1}\cdots w_{ij,M})$ now represent a set of colours. The terminology used to describe the SOM is summarised in Table \ref{tab:lexicon}.

To follow Kohonen's prescriptions -- i.e., to find the best-matching units and implement Eq.~(\ref{eq:w_update}) 
-- the distance between weights and galaxies is computed assuming a Euclidean metrics: 
\begin{equation}
   d_{ij} =  \sqrt{\sum_{m=1}^{M}\left(\mathcal{C}_m-w_{ij,m}\right)^2} \,,
   \label{eq:dist}
\end{equation}
where the given galaxy is defined by the colour vector $\hat{\mathcal{C}}=(\mathcal{C}_1, \mathcal{C}_2, ..., \mathcal{C}_{M})$.  The total $i\times j$ number of cells/weights is  chosen by the user (see Sect.~\ref{sec:som_exact}). 
As a result each SOM cell ``contains'' one or more galaxies from the training sample, whose colours are similar to the weight vector of that cell. 
The galaxy-cell association determined during the training phase is performed by the \texttt{python} software \textsc{SOMPY}\footnote{\url{https://github.com/sevamoo/SOMPY} }.  Before the iterative process, to start with weights that are already close to the galaxy distribution, each weight vector is initialised by setting its colours via principal component analysis \citep[PCA,][]{chatfieldetcollins80} of the training sample. 
A parallelism between these weights and  PCA  eigenvectors can help understanding the SOM: its weights can be thought as a set of characteristic SEDs  that describe the panchromatic space. However, the SOM has important differences from a PCA (see   Sect.~\ref{subsec:som_exact_spectra}).  In particular PCA is a linear hyper-surface defined by principle components, so it cannot fully describe a non-linear manifold, which is what we expect the galaxy colour space to be.

Once the SOM is trained, new galaxies can be mapped onto the grid by finding the nearest weight vector to each of them through Eq.~(\ref{eq:dist}). Moreover, the grid can be labelled \textit{a posteriori} by looking to another galaxy property not included in the parameter space $\mathcal{M}$.  For example, one can consider the redshift distribution of galaxies within a given cell and take their median $\langle z\rangle^\mathrm{cell}$ to label that cell. Such explicit labelling  gives our method a key advantage over supervised ML because we keep control of the relationship between  features (the broad-band colours) and labels (the redshift in the example above). This means that if we have a model of the bias or errors in our data we can directly account for that. However, it also means that an additional \textit{calibration} phase is required to make the SOM work as an effective galaxy estimator.   Indeed,  as we will highlight in the following, galaxies clustered together in the colour space also share other (physical) properties.

\section{Self-organising map of \textsc{Horizon-AGN} galaxies}
\label{sec:som_exact}

We apply the procedure summarised in Section~\ref{subsec:intro_som} using \textsc{Horizon-AGN} galaxies as a training sample. The considered features are their broad-band colours $u-B$,\,$B-V$,\,$V-r$,\,$r-i^+$,\,$i^+-z^{++}$,\,$z^{++}-Y$,\,$Y-J$,\,$J-H$,\,$H-K_\mathrm{s}$,\,$K_\mathrm{s}-[3.6]$,\,$[3.6]-[4.5]$.  Except for the Subaru intermediate-band filters, which are not included here, this is the same baseline used in Paper I. 
As the SOM projects that 11-dimensional space into a rectangular grid (likewise Fig.~\ref{fig:som_didact}) we can explore galaxy physical parameters as a function of their 2-D position, to see whether training objects located in the same cell have in common other properties besides their broad-band colours. It should be noticed that a galaxy  not detected in any of the filters poses a challenge to the SOM as one colour would be ill-defined. This is a common problem in ML methods that will be addressed in Sect.~\ref{sec:estimator}. 

 In the present Section we consider intrinsic colours (i.e., not affected by photometric noise) unless specified otherwise. The results discussed here are instrumental to show the fundamental properties of our method and its full potential in the case of an ``ideal'' survey. In Sect.~\ref{sec:real_calibr} we will address the impact of observational uncertainties. 

\subsection{Generating an \textit{ideal} SOM }
\label{subsec:som_exact_setup}

\begin{figure}
    \centering
    \includegraphics[width=0.9\columnwidth]{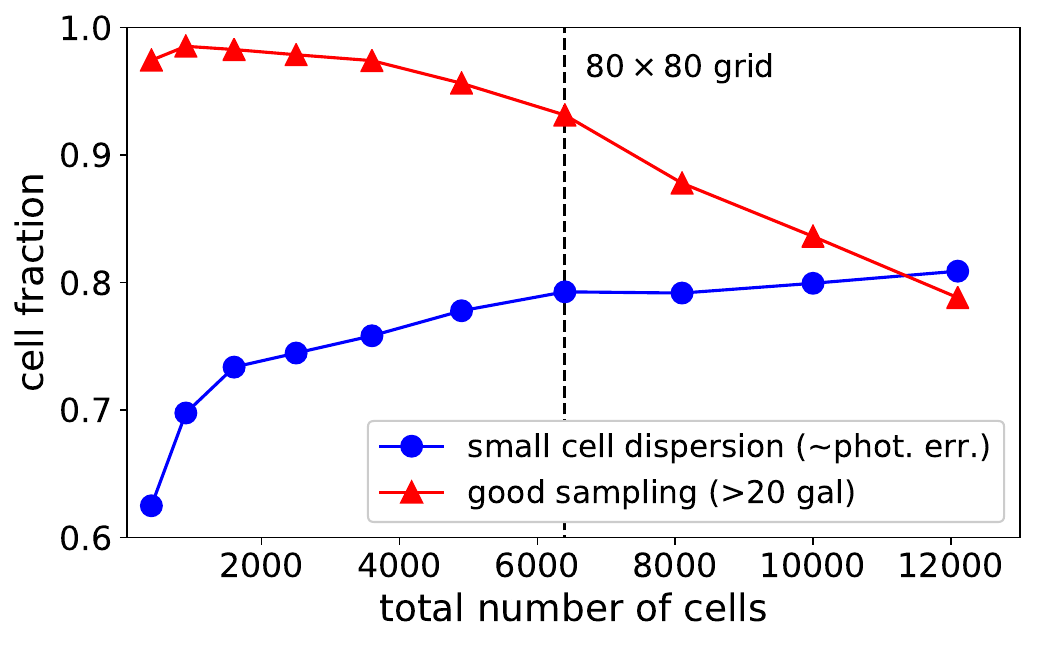}\\
    \includegraphics[width=0.9\columnwidth]{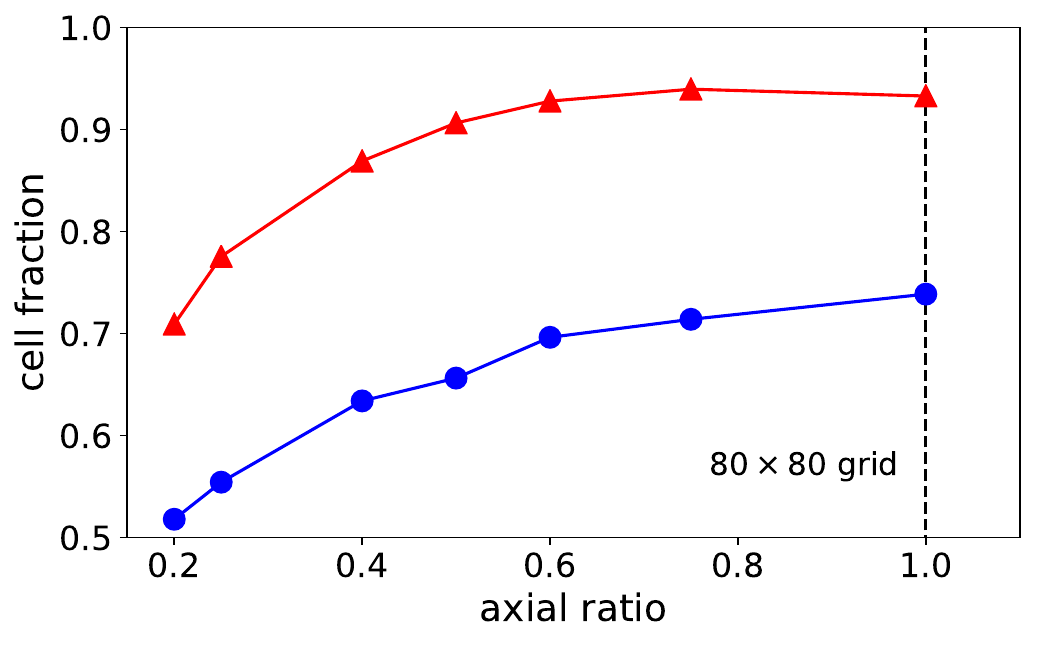}
    \caption{Tests to define size and shape of the SOM   grid to optimally describe the galaxy parameter space.  Different 2-D configurations are adopted, and each time the SOM algorithm is applied to the same \textsc{Horizon-AGN} sample (see Sect.~\ref{subsec:som_exact_setup}). The quality of each configuration is measured by the number of cells with a good sampling (i.e., including $>$20 galaxies, red triangles) and high clustering (cells where the average distance of galaxies from the weight is comparable with typical photometric errors, blue circles).  \textit{Upper panel:} Fraction of cells satisfying both quality criteria while increasing the size of the grid. Only square configurations are considered, starting from a $20\times20$ cells up to $110\times110$ (i.e., 12,000 cells in total). The $80\times80$ configuration  shows a good performance before the number of galaxies per cells starts to drop in larger grids.   
    \textit{Lower panel:} The two quality criteria this time are applied to rectangular grids with the same number of cells (6,400) but different aspect ratio.  As in the upper panel, red triangles show the fraction of cells populated with a conspicuous number of galaxies while blue circles show the fraction of cells in which galaxies are tightly clustered around the weight. 
    The $80\times80$ grid gives the best SOM.  
    }
    \label{fig:som_size}
\end{figure}

 Since we decided to adopt a rectangular lattice for the SOM the first step is to decide its size and axial ratio\footnote{
 Other 2-D configurations are possible, e.g.\ a lattice made by hexagonal cells or a spherical projection divided in  \textsc{healpix}  (see \citealp{CarrascoKind14} for a comparison).}. We set them by iteration, looking at changes in galaxy dispersion as a function of those two quantities (i.e., how tightly clustered are the galaxies associated to a given cell). We also check that the number of galaxies per cell is large enough to assure a good sampling in the various regions of the parameter space. 
 
 First we test the optimal SOM size. Starting from a $20\times20$ grid, we gradually increase the size by adding ten cells in both dimensions (i.e., maintaining a ``square'' configuration). Each time we train the SOM with the whole Horizon-AGN catalogue and measure i) the average distance of galaxies from their best-matching weight, ii) the number of galaxies associated to each weight.  
 The SOM converges fast with respect to i), so that in grids of $\gtrsim$6,000 cells most of the galaxies are tightly clustered in their cells, i.e., their distance from the weight in the colour space is smaller then the typical photometric errors in deep surveys like COSMOS (0.01$-$0.05 mag from optical to IR). On the other hand, the larger the grid of weights, the fewer the number of galaxies associated to each of them. With the  $90\times90$ configuration a significant area of the SOM starts to be under-sampled, with about 15 per cent of the cells defined by less than 20 galaxies each  (Fig.~\ref{fig:som_size}, upper panel). 
 Therefore we identify a slightly smaller size ($80\times80$ cells) as a good compromise between high resolution and sampling, also considering computational efficiency. 
 
 The next step is to define the best geometry for our SOM, namely the ratio between its axes. In the previous test we used only square grids with increasing number of cells, while now we fix the number of cells to $6,400$  and modify the aspect of the grid from 1:5 ratio to 1:1 (i.e., the $80\times80$ configuration) in eight steps. We describe again the quality of each SOM in terms of i) and ii), finding that the best configuration is that with 1:1 axis ratio (Fig.~\ref{fig:som_size}, lower panel). In a rectangular grid there are more galaxies not well represented (i.e., far from their best-matching weight) especially when the two sides have very different lengths. 
 
 In conclusion, the SOM we will use throughout is made by $80\times80$ cells. The result is specific for the \textsc{Horizon-AGN} parameter space: the optimal configuration for another galaxy sample may be different. 
 We notice nonetheless that the total number of cells is comparable to those used to describe the real COSMOS and CANDELS  data sets, respectively in \citet[][11,250 cells for galaxies up to $z\sim6$]{masters15} and \citet[][4,800 up to $z\sim4$]{hemmati19}. An alternate method, proposed by \citeauthor{hemmati19}, consists in increasing the grid size until the histogram of each weight vector component matches the distribution of the corresponding colour (see their figure 6). We calculate these histograms and find that indeed they converge when the grid dimension is $\geqslant80\times80$. 
 
 \begin{figure}
    \includegraphics[width=1.0\columnwidth]{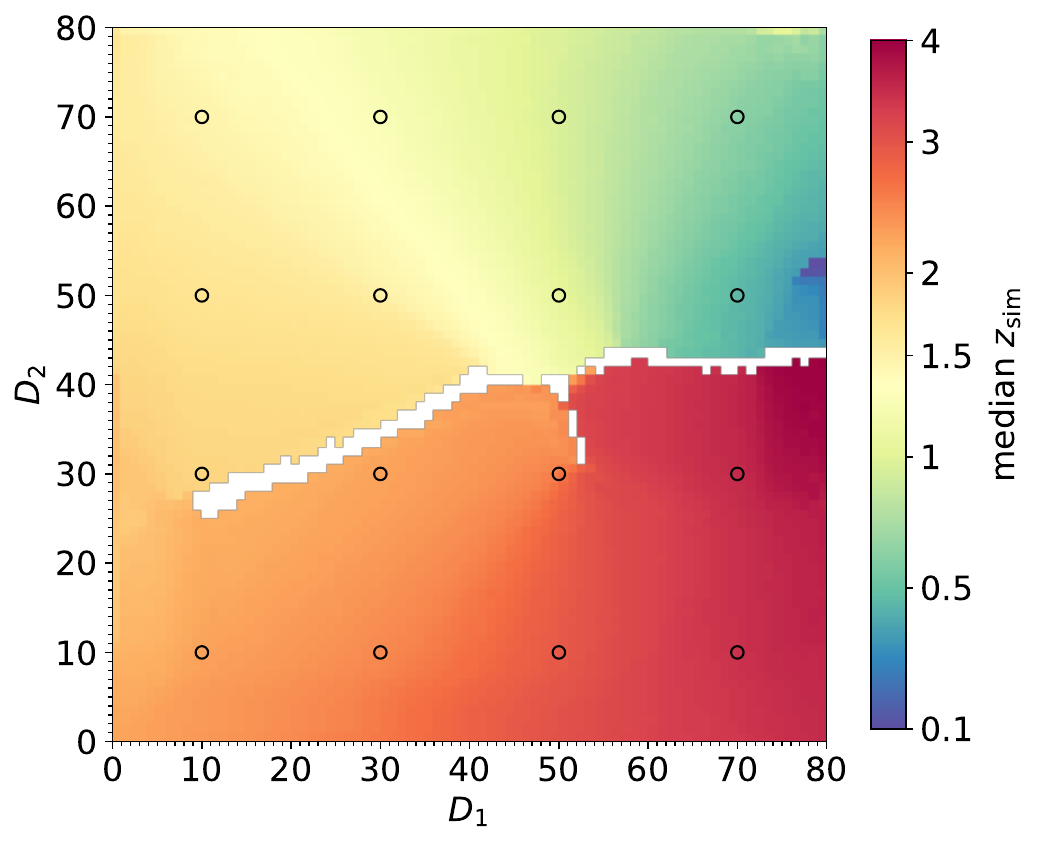} 
    \caption{SOM trained with a sample of \textsc{Horizon-AGN} galaxies using their intrinsic  colours (observer's frame). The grid is colour-coded according to the median intrinsic redshift of galaxies in the same cell ($\langle z_\mathrm{sim} \rangle^\mathrm{cell}$). Black empty circles identify cells whose galaxy spectra are analysed in Fig.~\ref{fig:stack_exact}. The white region is made by empty cells with no galaxy matching their weight. Despite  redshift evolution is overall smooth across the map, in the upper-right corner ($72<D_1<79,D_2=80$) a streak of  $\langle z_\mathrm{sim} \rangle^\mathrm{cell}\sim1.5$ cells show the small impact of border effects (see Sect.~\ref{subsec:som_exact_z}). }
    \label{fig:som_true_zsim}
\end{figure}

\subsection{Redshift calibration}
\label{subsec:som_exact_z}

  After the training phase, we can label each SOM cell according to a given property of the galaxies contained in it.  We compute the median redshift 
  $\langle z_\mathrm{sim}\rangle^\mathrm{cell}$  and the relative scatter $\sigma_z$ defined as $\sigma(z_\mathrm{sim}-\langle z_\mathrm{sim}\rangle^\mathrm{cell})$.
 The SOM of Horizon-AGN shows the same cell-$z$ relationship found by \citet{geach2012} in COSMOS, with a smooth redshift evolution as moving across the 2-D space (Fig.~\ref{fig:som_true_zsim}). The redshift scatter in every cell (not shown in the Figure) is particularly small, with $\sigma_z$ always between 0.05 and 0.1. The small scatter is even more remarkable when the $\sigma_z$ values are divided by the $1+z$ factor, shrinking to 0.01$-$0.03.  This is due to the algorithm's ability of clustering objects with very similar (observer's frame) colours, which correspond to similar redshift. One may expect some redshift interlopers -- i.e., objects with a significantly different $z_\mathrm{sim}$ from the rest of the cell --  due to SED degeneracies \citep{papovich01}.  However, we do not find this in the ideal case discussed here.   
 On the other hand, we do observe boundary effects produced by galaxies with extreme colours, which lie at the limits of the panchromatic manifold (see also the example in Fig.~\ref{fig:som_didact}). Those galaxies are pushed to the border of the grid, but with a negligible impact on the redshift distribution inside a cell (the redshift scatter remains modest: $\sigma_z\simeq0.1$).  

 In Fig.~\ref{fig:som_true_zsim} we also observe a long horizontal stripe of empty cells. No galaxy has been associated to their $\hat{w}$ during the training. The $\langle z_\mathrm{sim}\rangle^\mathrm{cell}$ labels explain the physical meaning of this empty region: it is a  ``caustic'' in the parameter space dividing $z\sim3$ galaxies  from those at lower redshift with similar colours. Since we are working with intrinsic photometry, their Lyman vs Balmer break degeneracy \citep[e.g.][]{stabenau08} is fully disentangled.

\subsection{High-resolution spectra in the SOM cells}
\label{subsec:som_exact_spectra}

\begin{figure*}
    \centering
    \includegraphics[width=0.8\textwidth]{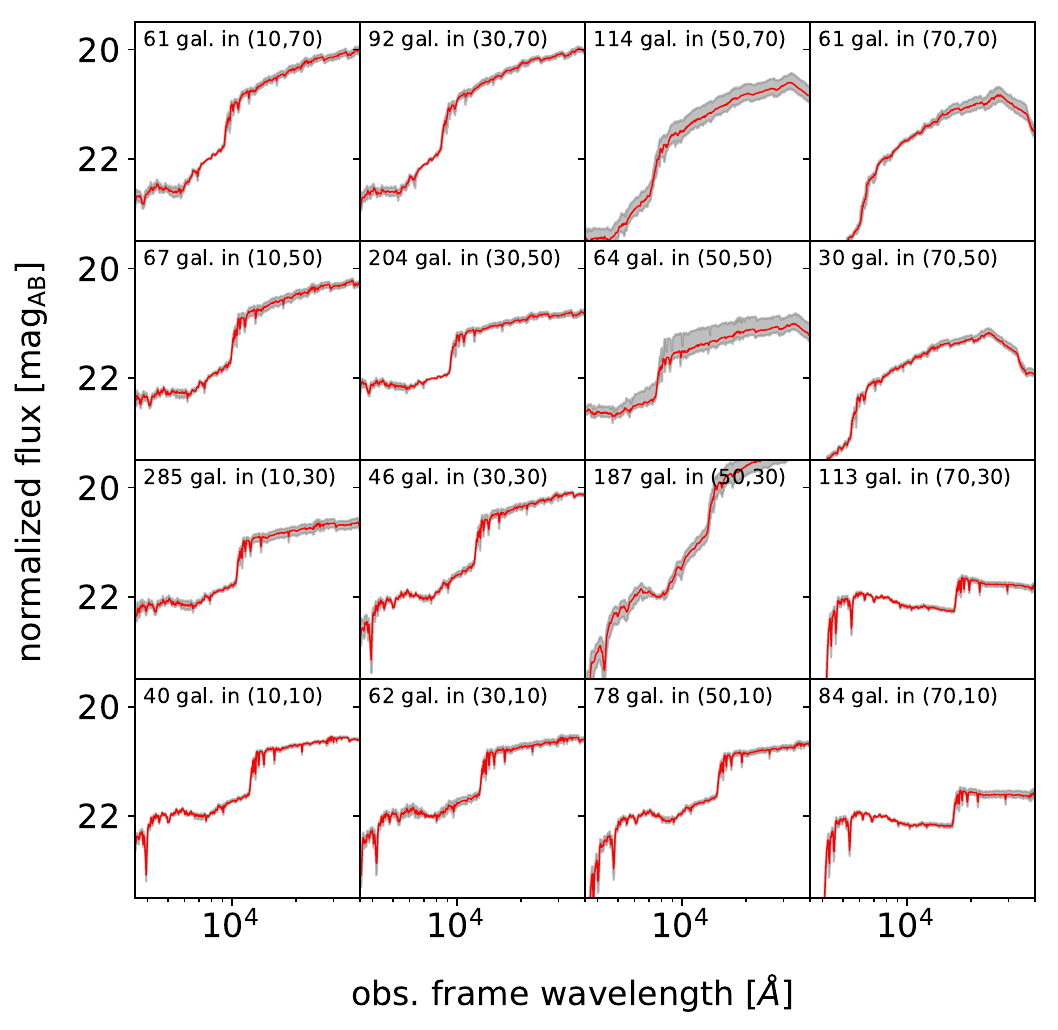}
    \caption{Median stacked spectrum (observer's frame, red line) of galaxies in a given cell with  relative inter-quartile dispersion (grey shaded area). Label at the top of each panel indicates number of galaxies and cell coordinates (see also empty circles in Fig.~\ref{fig:som_true_zsim}). Individual spectra are assumed to be ``observed'' at $z_\mathrm{sim}$ and then stacked, as one would do in a real survey. They have also been re-normalised to the arbitrary  $i^+$-band flux of 22\,mag. }
    \label{fig:stack_exact}
\end{figure*}

 In the {\sc Horizon-AGN} virtual observatory the broad-band colours are integrated from high-resolution BC03 spectra, which are built accounting for complex SFH and chemical enrichment (Sect.~\ref{subsec:intro_simu}).  
 
 In the SOM we stack galaxy spectra from the same cell, re-normalising them to a fixed $i^+$ flux to ease the comparison. Stacking is performed in the observer's reference frame: spectra are redshifted to their  $z_\mathrm{sim}$ before adding them, to be consistent with the analogue procedure one would implement in a real survey. 
 
 The fact that the colour-based SOM can efficiently map redshifts (Sect.~\ref{subsec:som_exact_z}) does not necessarily imply that it performs as well with spectral features. 
 Given the small amount of spectroscopic information in real data sets, previous work has only proven that broad-band SEDs are well clustered within the grid, with the exception of \citet{rahmani18} analysing 142 spectra at $0.5<z<1$ and \citet{hemmati19} showing a handful of  $z\sim1$ spectra that have  similar shape and are also clustered in nearby cells. 
  We find that this is actually the case for the whole \textsc{Horizon-AGN} sample:  galaxy spectra are in excellent agreement in most of the cells (see a few examples in Fig.~\ref{fig:stack_exact}). Inspecting the regions close to the redshift caustic we find more  dispersion, mainly because the median stacking is performed in observer's frame and therefore it is affected by the difference between  individual redshifts. We check that spectral shapes are even more similar if the comparison is made in rest frame, removing the $\sigma_z$ scatter.  
 
 In addition to the examples shown in Fig.~\ref{fig:stack_exact} we perform the same stacking analysis in 225 distinct cells evenly distributed across the grid\footnote{Those cells have coordinates $D_1=5i$ and $D_2=5j$, where $i$ and $j$ are integers ranging from 1 to 15.}, finding that their average inter-quartile dispersion is always $\lesssim$15 per cent ($<8$ percent in half of the cells probed). 
 It is worth emphasising the differences between this result and what a PCA classification would give.  
 First of all, PCA provides a basis of eigenvector to be linearly combined, whereas the SOM works also with non-linear  transformations. Each weight in the SOM  has a clear physical meaning by itself, i.e., it describes a galaxy \textit{phenotype} in the observed frame \citep[][]{sanchez&bernstein19}. On the other hand, in a PCA  classification, the 2$-$4  eigenspectra that usually have the most discriminating power are difficult to interpret. They can be combined to reproduce actual galaxy features, but classes for those resulting spectra are not inherently provided by the PCA and human intervention is required \citep[e.g., defining meaningful regions in a Karhunen-Lo\`{e}ve diagram,][]{marchetti2013}.

 The degree of similarity of the simulated spectra within a given SOM cell also depends on the complexity of their features. Spectra used in this work are extracted from the {\sc Horizon-AGN} and their realism and complexity are constrained by those of the simulation. Modelling galaxy evolution on cosmological scales is inevitably done at expenses of  resolution. Because {\sc Horizon-AGN} maximum spatial resolution is at best 1\,pkpc, the impact on the inter-stellar medium (ISM) of any process occurring at a smaller scale is averaged through sub-grid recipes. These recipes have been iteratively improved  in order to reproduce as well as possible the statistical distribution of integrated galaxy properties throughout cosmic time \citep[in \textsc{Horizon-AGN} such a progress can be tracked through][]{dubois12,dubois14,park19}. However, they might fail to some extent at reproducing ISM in-homogeneity and clumpiness. As a result, we expect the  SFH of our simulated galaxies to be  smoother (and their  spectra less diverse) than the real ones.  As a consequence, the dispersion that we measured within a SOM cell must be considered as a lower limit. A more extended discussion about caveats in our modelling is provided in Appendix~A1.

 \subsection{Exploring other physical parameters in the SOM}
\label{subsec:som_exact_phys}

\begin{figure*}
    \centering
    \includegraphics[width=\textwidth]{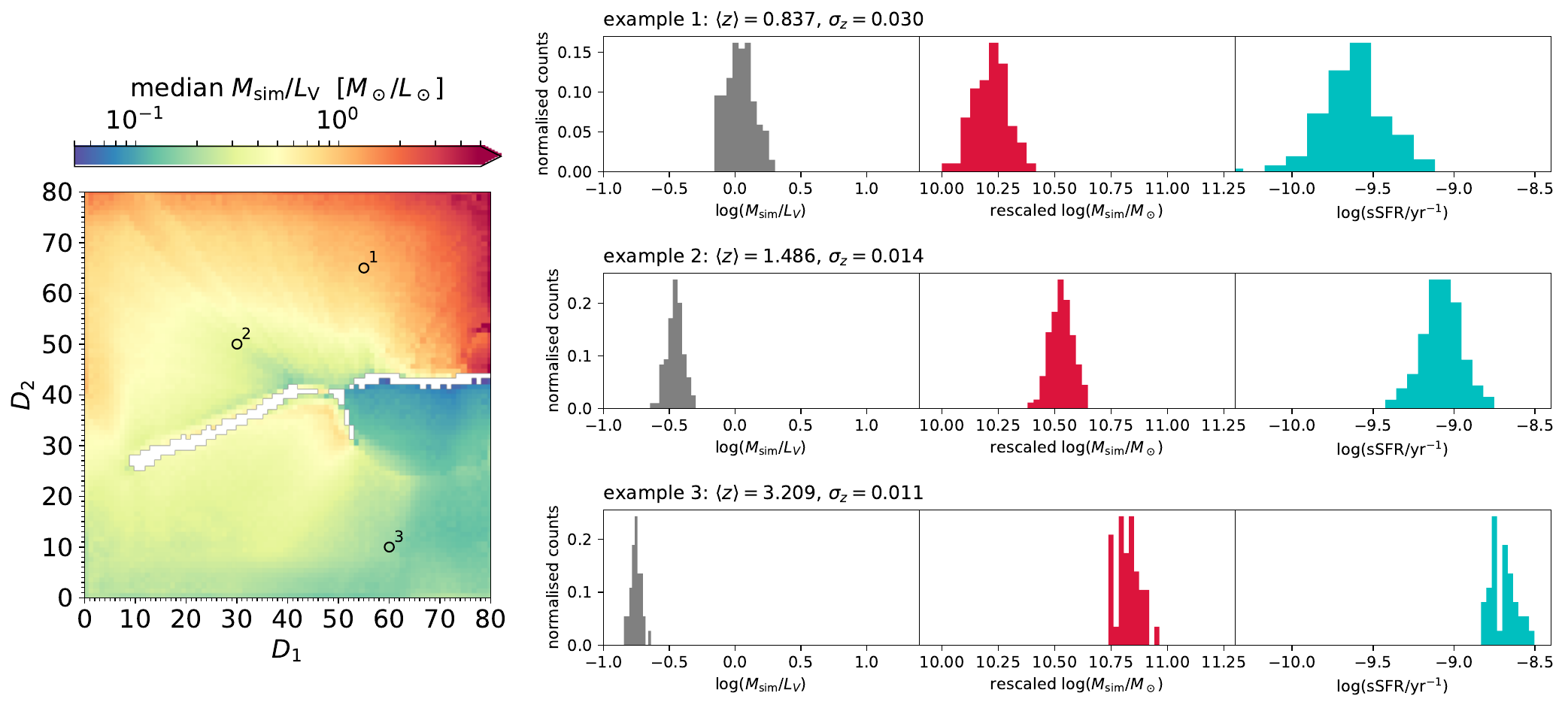}
    \caption{The SOM trained with the intrinsic colours from the \textsc{Horizon-AGN} catalogue is able to group galaxies with similar physical properties together in the same cell. \textit{Left:} The SOM grid (same as Fig.~\ref{fig:som_true_zsim}) is now labelled according to the median mass-to-light ratio in each cell. Empty circles mark three (randomly extracted) cells used as examples to show the tight correlation between position in the grid and physical properties. \textit{Right:} Each row corresponds to one of the example cells marked in the left panel and shows the logarithmic distribution of $M_\mathrm{sim}/L_V$ (grey histogram), $M_\mathrm{sim}$ (red histogram), and sSFR (cyan histogram) in that cell. In the case of the  $M_\mathrm{sim}$ distribution, individual values have been normalised to a common $i^+$-band flux (corresponding to 22\,mag) to show that the dispersion in a given cell is mainly due to the fact that the SOM, being trained with colours only, is not informed about the normalisation of the SED. }
    \label{fig:som_true_mtol}
\end{figure*}

  The remarkable similarity between spectra in the same cell suggests that those galaxies went through a similar evolutionary path, resulting e.g.\ in the same mass-to-light ratio at the redshift of observation. 
  \textsc{Horizon-AGN} provides us with galaxy physical parameters such as luminosity ($L$), stellar mass ($M_\mathrm{sim}$), and star formation rate (SFR$_\mathrm{sim}$). Thus, we can use our simulation to test whether the objects that are grouped toegther by the SOM have other properties in common besides their redshift. Regarding SFR$_\mathrm{sim}$, this is defined by averaging galaxy SFH over the last 100\,Myr, an interval comparable to the timescale of SFR indicators widely used in the literature \citep{kennicutt&evans12}. 
  
  We calibrate the SOM to show the typical mass-to-light ratio per cell  (Fig.~\ref{fig:som_true_mtol}, left panel). The procedure is similar to the redshift calibration in Sect.~\ref{subsec:som_exact_z}, this time computing the median $\langle M_\mathrm{sim}/L_V \rangle^\mathrm{cell}$ with $L_V$ being the luminosity in the rest-frame $V$ band. In this way we can visualise the variation of this quantity across the grid, which is also an evolution across redshift: more mature galaxies with larger $M/L$ occupy cells with lower $\langle z_\mathrm{sim}\rangle^\mathrm{cell}$ (cf.\ Fig.~\ref{fig:som_true_zsim}). We also aim at verifying that the scatter inside a given cell is small by  
  calculating the difference between the 84th and 16th percentile in the logarithmic 
  $M_\mathrm{sim}/L_V$ distribution. We find that in most of the cells this is smaller than 0.2\,dex. A few examples of the tight correlation between position in the grid and $M/L$ are shown in the histograms of Fig.~\ref{fig:som_true_mtol}.  
  
\begin{figure}
    \centering
    \includegraphics[width=\columnwidth]{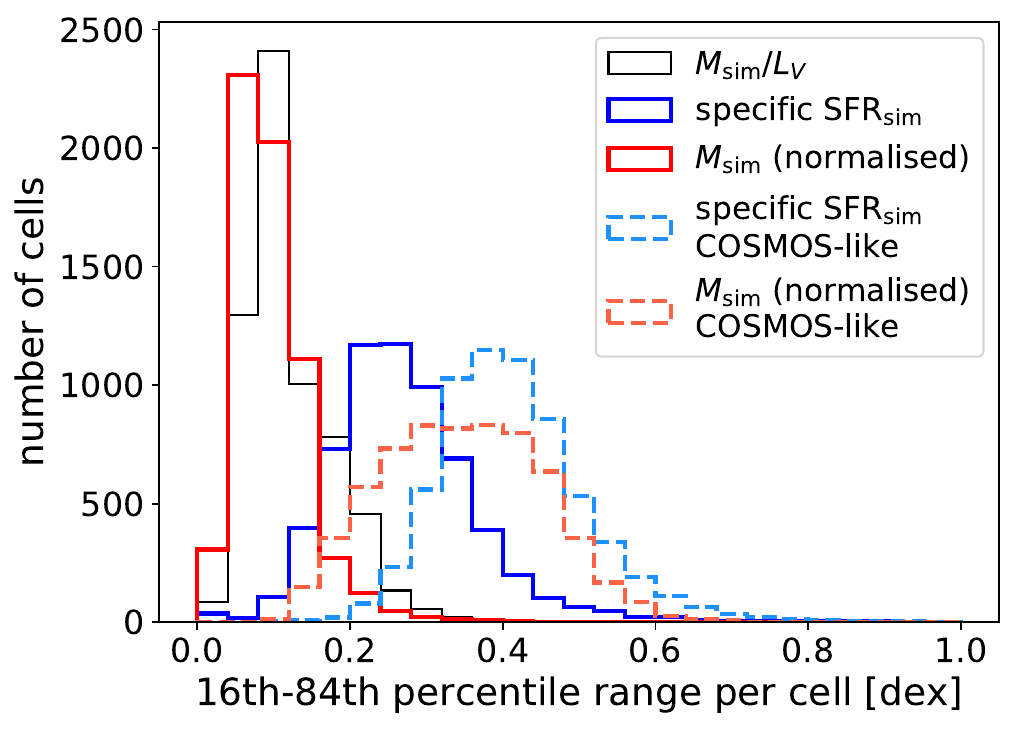}
    \caption{Logarithmic dispersion of various  physical properties in each cell of the SOM, defining such a scatter as the difference between 84th and 16th percentile of the logarithmic mass-to-light ratio (thin black histogram), sSFR (blue), stellar mass (red). For sSFR and stellar mass, solid lines show the results when intrinsic colours are used (see Sect.~\ref{subsec:som_exact_phys}) while dashed-line histograms are the case in which galaxy photometry is perturbed with COSMOS-like errors (Sect.~\ref{subsec:som_noise}). Since the SOM is trained only with colours, an SED normalisation has to be applied for stellar masses;  we used  the $i^+$-band fluxes to this purpose, re-scaling galaxies to $i^+=22$ as in Fig.~\ref{fig:stack_exact}.    }
    \label{fig:som_true_1684hist}
\end{figure}

  A similar trend can also be observed regarding stellar mass after a scaling factor is applied. This factor is required since we do not train the SOM  with information about SED normalisation, as instead other authors do with different ML methods \citep{bonjean19}\footnote{
  We prefer  working in a pure colour space because an additional dimension with a different dynamical range may not be properly weighed with respect to the others.}.
  This is the same procedure used in Fig.~\ref{fig:stack_exact} to compare galaxy spectra, which have similar shapes but different magnitude.  
   Therefore, to analyse the intrinsic $M_\mathrm{sim}$ scatter within a given cell, we first normalise each galaxy to a reference point of $i^+=22$\,mag. 
   In principle this should be done in rest frame, given the fundamental $M/L$ correlation, but since spectra in the same cell are at about the same redshift one can use apparent magnitudes instead. 
   After such a re-scaling, the $\log(M_\mathrm{sim}/M_\odot)$ dispersion is smaller than 0.2\,dex in most of the cells (see three examples in Fig.~\ref{fig:som_true_mtol}, and also  Fig.~\ref{fig:som_true_1684hist}). 
   
   The specific SFR (sSFR$\equiv$SFR$_\mathrm{sim}/M_\mathrm{sim}$) does not need such a normalisation as it is expected to be directly related to $M/L$. However the scatter is larger than $M/L$ and also stellar mass. In fact, the latter is an integrated quantity strongly connected to the global evolutionary path of the galaxy, whereas the sSFR depends on recent fluctuations in the mass assembly history. The 0.2$-$0.5\,dex dispersion in the SOM (Fig.~\ref{fig:som_true_mtol} and \ref{fig:som_true_1684hist}) reflects such a  stochasticity; sSFR is also more sensitive than stellar mass to different levels of dust attenuation. 
   
   These tight correlations are obtained with a SOM trained with intrinsic colours: larger scatter is expected when galaxy photometry is affected by observational uncertainties. 
   The results shown in this section can be considered as an ``asymptotic'' limit represented by an ideal galaxy survey with infinite $S/N$. 
   This ideal example can be approximated by the brightest galaxies in an ``ultra-deep'' survey. In that case the $S/N$ should be high enough to enable the SOM 
   to classify galaxies not only with respect to their redshift, but also stellar mass and SFR. Such a clustering ability is that cornerstone of our original technique to recover stellar mass and SFR. 
   In principle, knowing the stellar mass of a ``calibration'' object ($M_\mathrm{cal}$, which can be independently obtained via template fitting) one could get a fairly precise $M$ estimate for other galaxies mapped into the same cell: it would suffice to re-scale $M_\mathrm{cal}$ by the flux ratio between the known galaxy and the target one: $M\simeq M_\mathrm{cal}\times(f/f_\mathrm{cal})$.   
   As suggested by the examples in Fig.~\ref{fig:som_true_mtol} in the best-case scenario the uncertainty of such an estimate would be comparable to the typical mass errors of template fitting codes \citep[$\lesssim0.3$\,dex, see e.g.][]{davidzon17} with significant improvement in computational speed \citep[about 10$^6$ times faster according to][]{hemmati19b}. 
   
   The same argument used for stellar mass applies to SFR, which is not shown in Fig.~\ref{fig:som_true_mtol} as it will be thoroughly 
   discussed in the following. A flux re-normalisation will be necessary as in the case of stellar mass. 
   In the present analysis we use the $i^+$ band for the flux scaling factor since  it has one of the deepest sensitivity limits. The $K_\mathrm{s}$ band is a better proxy for stellar mass but in COSMOS2015 (and therefore in our COSMOS-like sample) its $3\sigma$ limit $\sim$1.5 mag shallower than $i^+$ and would result in a larger scatter of the results (but still preserving the  properties of the SOM, see Appendix~A2).

\section{A novel estimator of galaxy SFR}
\label{sec:estimator}

 We concluded Sect.~\ref{sec:som_exact} by suggesting that the SOM can be used to empirically recover galaxy physical parameters in the same framework used in \citet{masters15} for photometric redshift computation.
 In this section we discuss how to calibrate the SOM in order to derive SFRs from broad-band colours in a fast but accurate way.   
 We focus on SFR since standard SED fitting shows its limit when deriving this quantity (see Paper I). 
 One can obtain robust SFR estimates based on IR imaging or spectroscopy, but this is generally possible for a small fraction of galaxies (which indeed we can use as a  calibration sample for our method). 
 A SOM-based estimator for stellar masses should also be feasible, but it would likely be based on template fitting, at least for calibration purposes, given the current state of the art in estimating stellar mass\footnote{
 Other options are available in some case, e.g.\ dynamical masses from spectroscopy \citep{courteau14}. }.  For template based approaches, we refer the reader to \citet{hemmati19b} for an overview of how to effectively use the SOM. Here, our goal is to empirically calibrate properties based on direct indicators rather than  depending on synthetic SED libraries. 
 
 With this goal in mind we devise a method that can be applied to real data, requiring some adjustment to the SOM. 
 We replace the ideal (noiseless)  photometry used in the previous section with a catalogue that mimics the COSMOS survey, including errors and selection functions.  In other words, apparent magnitudes and colours of \textsc{Horizon-AGN} galaxies are now perturbed with observational-like errors and selection effects. As a consequence we use a different training sample, selecting only galaxies above a given $S/N$ threshold. Thus, the observational-like training sample does not include objects that are not detected in some band.  In  Sect.~\ref{subsec:som_noise} we show that, after the $S/N$ selection, photometric errors do not impair the relationship between SOM cells and galaxy properties shown above. 
 
 In Sect.~\ref{subsec:measurements} we explain the details of our methods. In that context we modify the way to label SOM cells. In fact, in previous sections we made calibrated versions of the SOM by labelling its cells with median values of either redshift or other physical parameters, i.e., under the assumption of knowing them for the whole sample. Hereafter we assume to know the SFR of a small subset of 6,400 galaxies (i.e., one object per cell) and use them to label the grid. At the moment we do not make particular assumptions on how such a calibration sample is built; this is discussed later in Sect.~\ref{sec:real_calibr} where we also compare to SFR estimates from template fitting.

\subsection{The COSMOS-like SOM}
\label{subsec:som_noise}

 The observational uncertainties to perturb the \textsc{Horizon-AGN} photometry are statistical errors affecting apparent magnitudes so that our mock galaxy catalogue resembles the quality of COSMOS data (see Paper I).  For this reason in the following we refer to the (noisy) \textsc{Horizon-AGN} sample also with the term ``COSMOS-like'', in contrast to the previous version with intrinsic photometry (Sect.~\ref{sec:som_exact}). We do not model confusion noise and contamination by saturated stars; this kind of issues shall be addressed in future work after providing the \textsc{Horizon-AGN} virtual observatory with simulated images. 

 The training phase of the implemented algorithm does not account for cases of non-detection (e.g., when the ``observed'' flux is smaller than the flux error). Therefore, while working with perturbed photometry, we limit the analysis to a galaxy sub-sample with  $S/N>1.5$ in each broad-band filter. 
 A statistically correct treatment of lower and upper limits in the  input colours would require an improved SOM algorithm that is beyond the goal of the present work. We note that also template fitting codes often neglect such a treatment \citep[as highlighted in][]{sawicki12}. 
 The $S/N$ pre-selection will restrict the analysis to  $z\lesssim 3.5$ because galaxies at higher redshift are $u$-band drop-outs \citep{Steidel:1996p9040} with $S/N\ll1$ in that filter. Given the sensitivity of our catalogue the $S/N$ threshold roughly corresponds to a flux-limited survey with a cut at $i^+<25$  (see Fig.~B1 in the On-line Supplementary Materials). 
  Besides the removal of $z>3.5$ galaxies from the  sample there are other caveats in the $S/N$ selection, which are listed in Appendix~B1. None of them affects the analysis between $z=0.2$ and  $z\sim3$, but there is a ``boundary effect'' at the lowest and highest redshifts of the range  (see below).

 \begin{figure}
    \centering
    \includegraphics[width=0.9\columnwidth]{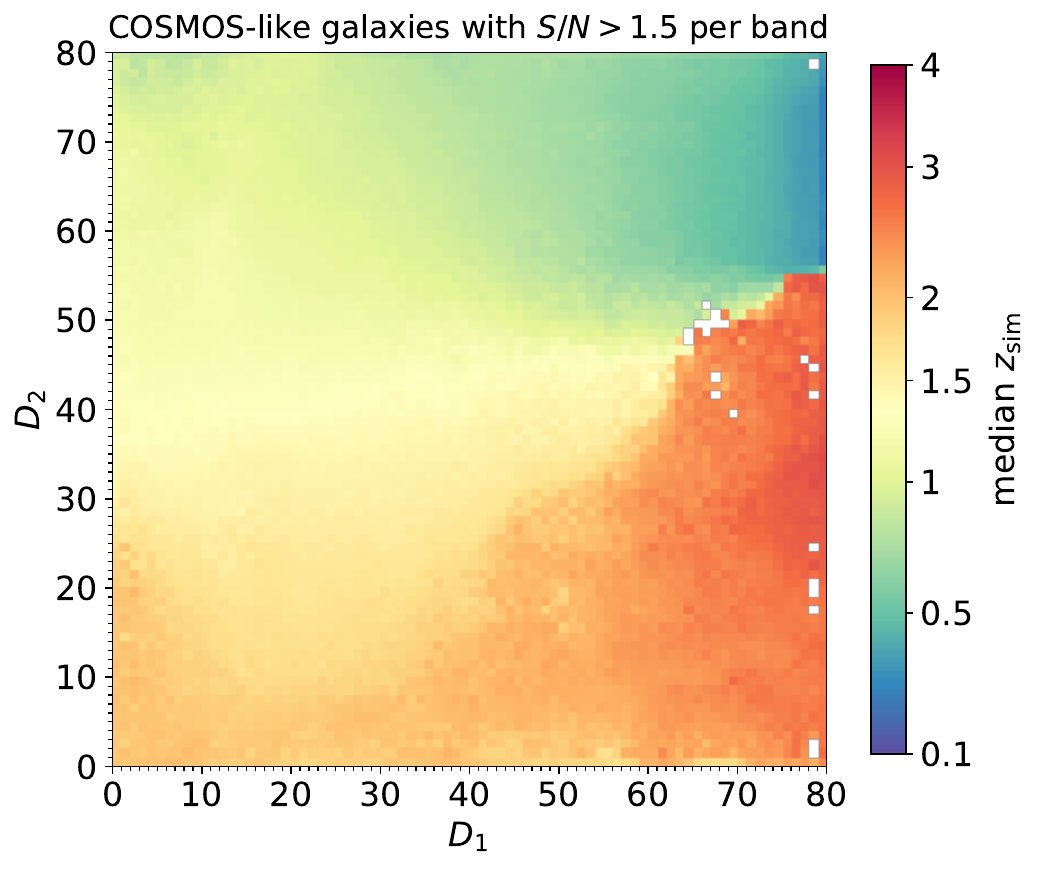}\\
    \includegraphics[width=0.9\columnwidth]{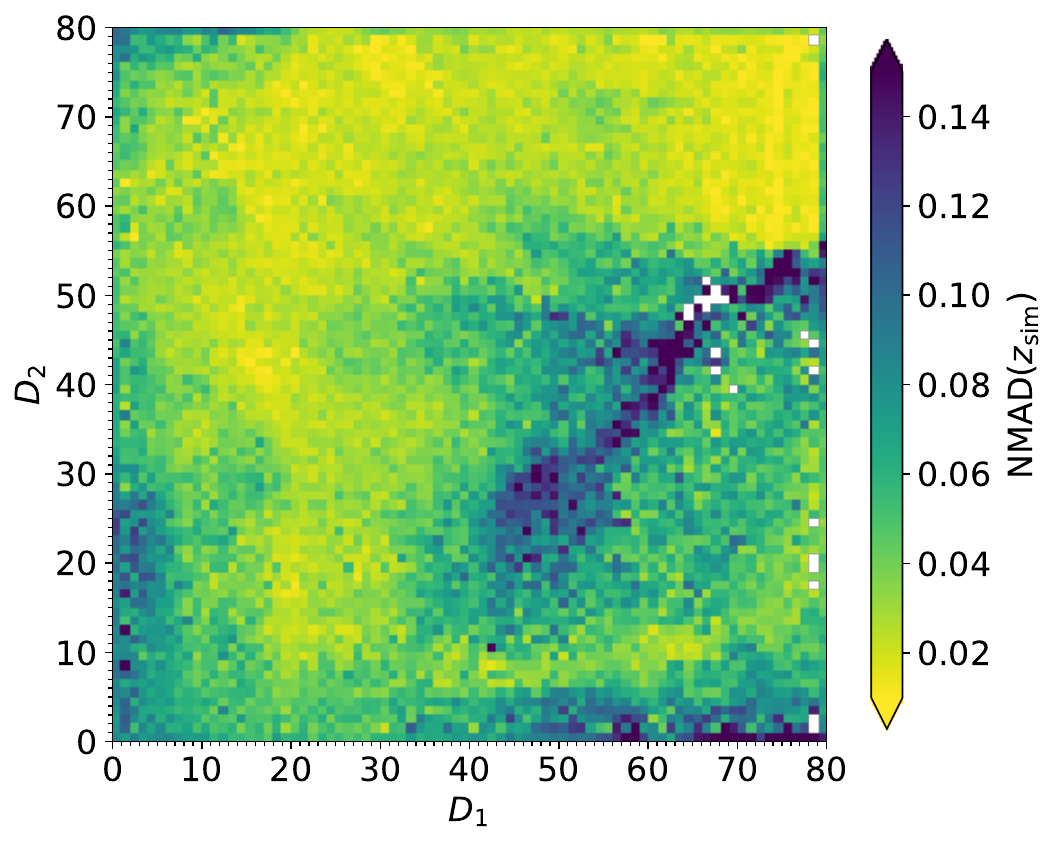}\\
    \includegraphics[width=0.9\columnwidth]{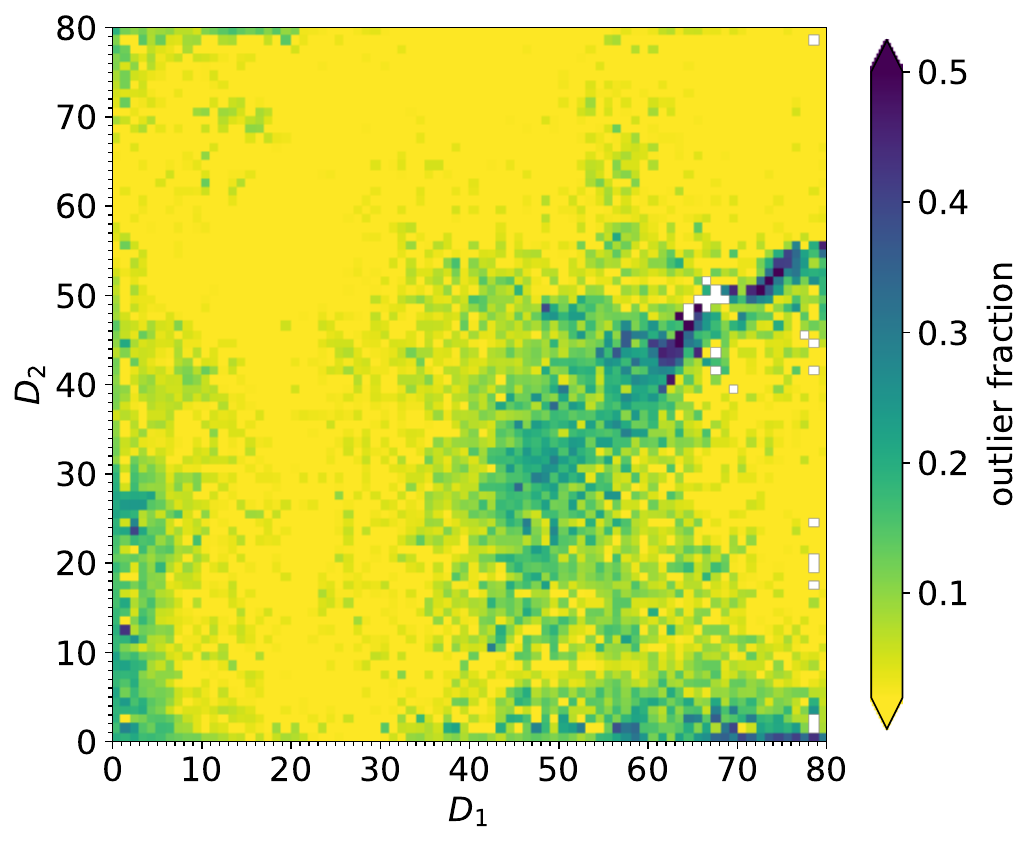}
    \caption{SOM algorithm applied to Horizon-AGN colour space, including photometric errors. A sub-sample of galaxies with robust colours ($S/N>2$ in all the photometirc pass-bands) is used to train and label the SOM. \textit{Upper panel:} Cells are colour-coded according to their median redshift (compare to Fig.~\ref{fig:som_true_zsim}). The redshift range is now $0<z<3.2$ because of the $S/N$ limit imposed in the $u$ band (see text). 
    For the same reason 13 cells sparsely distributed across the grid are empty (coloured in white). Other 9 empty cells identify the redshift ``caustic'', the rest of which is filled by scattered objects.  \textit{Middle panel:} normalised median absolute deviation (NMAD). \textit{Lower panel:} fraction of redshift outliers in each cell. }    \label{fig:som_noise_zsim}
\end{figure}

 After this preliminary test we use the perturbed photometry of \textsc{Horizon-AGN} galaxies with $S/N>1.5$ to produce a new SOM; the multi-dimensional space is the same  as in Sect.~\ref{sec:som_exact} (11 broad-band colours).   
 Figure~\ref{fig:som_noise_zsim} shows the resulting redshift map limited to $z_\mathrm{sim}\lesssim3.5$. The  redshift evolution across the $80\times80$ grid is similar to the ideal SOM, although some details are smeared out because of photometric errors. For example the gap between low- and high-$z$ regions is now filled by scattered galaxies  (cf.\  Fig.~\ref{fig:som_true_zsim}).    
 We quantify redshift dispersion in each cell through the normalised median absolute deviation \citep[NMAD,][]{hoaglin1983} and the outlier fraction. The former is defined as $1.48\times\mathrm{median}( \vert \Delta z\vert)/(1+ \langle z \rangle^\mathrm{cell})$, where $\Delta z \equiv z_\mathrm{sim} - \langle z  \rangle^\mathrm{cell}$ (Fig.~\ref{fig:som_noise_zsim},  middle panel). The latter is the fraction of objects in a given cell having 
 $\vert \Delta z \vert/(1+ \langle z \rangle^\mathrm{cell})>0.15$ (Fig.~\ref{fig:som_noise_zsim}, lower panel). These metrics are similar to those used to describe the quality of photometric redshifts estimated via template fitting \citep[$z_\mathrm{phot}$, see e.g.][]{ilbert13}. 
 The NMAD is on average $<0.047$, with only 343 cells above 0.1 (mainly in the formerly empty region of the $z$ caustic). The outlier fraction is overall small, being less than 10 per cent in 5,228 cells (namely 82 per cent of the grid).
 On the other hand systematics at the borders are more evident than before.  
 
 Besides redshift, \textsc{Horizon-AGN} galaxies remain well clustered also with respect to $M$ and sSFR  (see Fig.~\ref{fig:som_true_1684hist}).  With perturbed photometry, the sSFR$_\mathrm{sim}$ is slightly less constrained than in the ideal SOM: e.g., now there are no cells with scatter $<$0.2\,dex. On the other hand  only $\sim$10 per cent of the cells exceed 0.5\,dex dispersion in sSFR. These are cells hosting low-sSFR galaxies.  The results are comparable to other classification methods  \citep[e.g.\ the $\mathrm{NUV}-r$ vs $r-K$ diagram,][]{arnouts13}. 
 The typical scatter in $\log(M_\mathrm{sim}/M_\odot)$  within one cell is 0.3$-$0.4\,dex, much larger than before  because not only the input colours but also the $i^+$-band re-scaling now is done with perturbed fluxes. However, this is of the same order of $M$ statistical errors in observed galaxies \citep[e.g.,][]{davidzon17}, a further indication that the SOM estimator can work also in the observed universe.

\subsection{Galaxy redshift and SFR measurements}
\label{subsec:measurements}

 Encouraged by the previous tests, we proceed in the implementation of the SOM estimator. 
 First of all we need a reference sample to label the SOM grid with both redshift and SFR values. Therefore we assume to ``observe'' one galaxy per cell to obtain an estimate of their redshift and SFR. 
 These galaxies belong to the \textit{calibration sample} and their ``measured'' properties are dubbed $z_\mathrm{cal}$ and SFR$_\mathrm{cal}$. For the moment we do not make stringent requirements about how  $z_\mathrm{cal}$ and SFR$_\mathrm{cal}$ are measured: they may come e.g.\ from a spectroscopic survey, but not necessarily. We only make the assumptions that these are \textit{bona fide} galaxies with reliable $z$ and SFR, and they cover the entire $80\times80$ grid.    
 Each calibration galaxy is randomly targeted among those in the given cell, with a sampling rate of one target per cell. For sake of simplicity, we do not model observational uncertainty so the $z_\mathrm{cal}$ and SFR$_\mathrm{cal}$ values correspond to $z_\mathrm{sim}$ and SFR$_\mathrm{sim}$ of the given galaxy.  In Sect.~\ref{subsec:bona_fide} we will discuss which kind of survey might provide such a calibration sample, modifying $z_\mathrm{cal}$ and SFR$_\mathrm{cal}$ accordingly. 
 
 The other galaxies in the SOM, not used for the calibration, will get an estimate of redshift and SFR from the SOM through the procedure described here. The method takes into account not only the best-matching cell in which galaxies lie but also the nearby ones. This choice is motivated by the impact of colour uncertainties: even though the SOM training phase places any COSMOS-like galaxy is placed into its best-matching cell, the colours of that galaxy are still compatible (within error bars) with the weights of other cells. There is a non-negligible probability that one them, in absence of observational errors, would be the true best-matching cell for the given galaxy. 
 
 For each entry of the mock catalogue our algorithm includes the following steps:  
 \begin{enumerate}
     \item\label{step1} consider $N_\mathrm{c}$ cells: the best-matching unit in which the \textsc{Horizon-AGN} galaxy reside and the nearest $N_\mathrm{c}-1$ cells; 
     \item\label{step2} calculate the distance between the galaxy and each of those cells, with a modified version of Eq.~(\ref{eq:dist}) that takes into account photometric errors:
     \begin{equation}
           d_{i} =  \sqrt{\sum_{n}\left(\mathcal{C}_n-w_{i,n}\right)^2 / \Delta \mathcal{C}_n^2} \,, \label{eq:dist_chi}
     \end{equation}
     where $\Delta \mathcal{C}_n$ is the 1$\sigma$ uncertainty for the n-th colour and $i$ is one of the $N_\mathrm{c}$ cells; 
     \item\label{step3} take the $z_\mathrm{cal}$ and  SFR$_\mathrm{cal}$ labels of the $N_\mathrm{c}$ cells;     
     \item\label{step4} compute their distance-weighted mean $z_\mathrm{SOM}$ and SFR$_\mathrm{SOM}$. 
 \end{enumerate}
 In particular the resulting SFR is re-normalised (as done for stellar masses in Sect.~\ref{subsec:som_exact_phys}):  
     \begin{equation}
          \mathrm{SFR}_\mathrm{SOM} = \frac{ \sum_{i \in N_\mathrm{nc}} (\mathrm{SFR}_\mathrm{cal}/d_i^2)  \times(f/f_\mathrm{cal,i}) }{ 
          \sum_{i \in N_\mathrm{c}}  1/d_i^2 } \,, 
          \label{eq:sfr_som}
     \end{equation}
    where $f/f_\mathrm{cal,i}$ is the flux ratio in a reference band (we choose $i^+$) between the given photometric galaxy and the \textit{bona fide} one that labels the $i$-th cell. The square distance from the $i$-th cell is also used in the weighted mean to compute $z_\mathrm{SOM}$. 
    
\begin{figure}
     \centering
     \includegraphics[width=0.99\columnwidth]{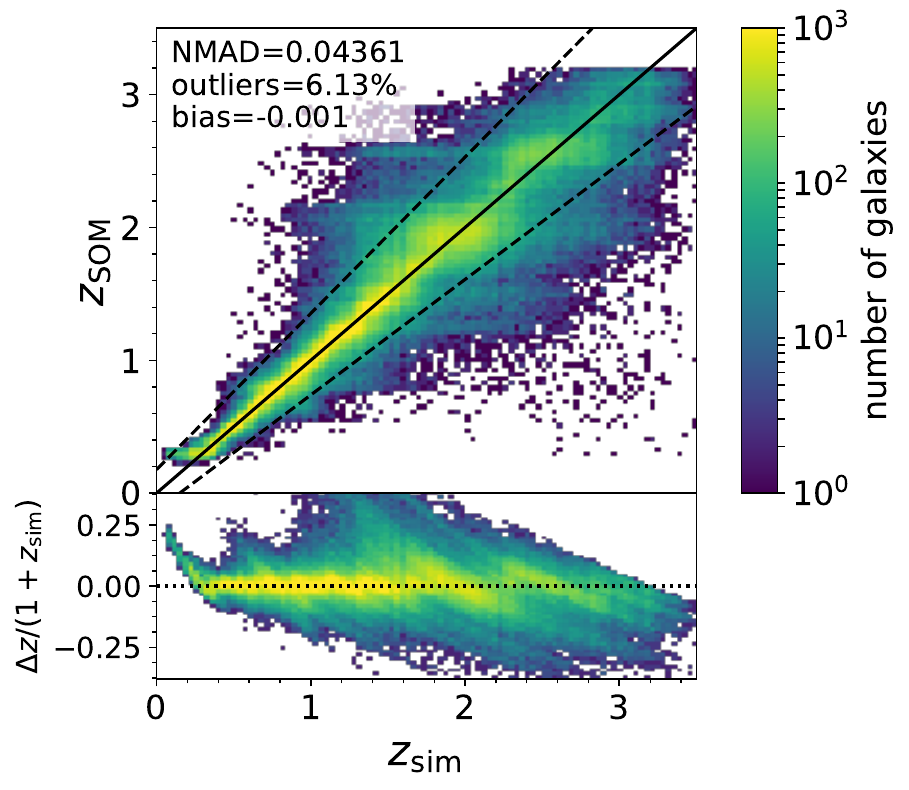}\\
     \includegraphics[width=0.99\columnwidth]{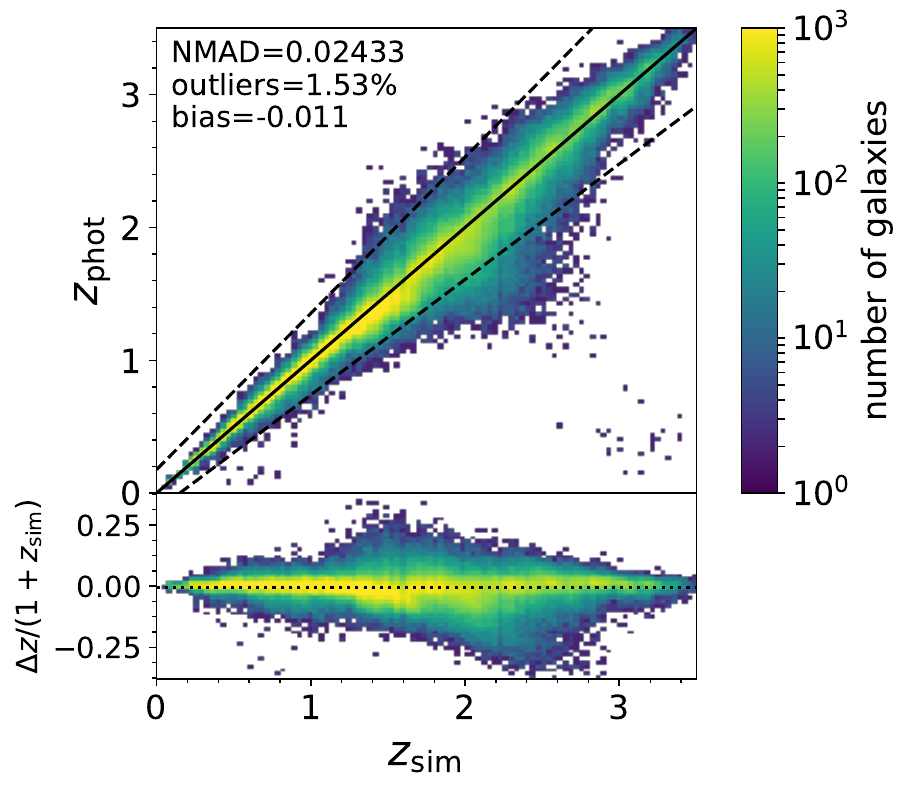}
     \caption{\text{Upper panel:} photometric redshifts derived from the  SOM estimator ($z_\mathrm{SOM}$) compared to the intrinsic redshifts ($z_\mathrm{sim}$), for the 371,168 COSMOS-like training sample not used for calibration. A solid line shows  the 1:1 bisector while dashed lines mark the $\pm0.15(1+z)$ threshold used to compute the outlier fraction. Metrics describing the quality of results are quoted in the upper-left corner.  The bottom of the figure shows the scatter $\Delta z/(1+z_\mathrm{sim})$ as a function of redshift.
     \text{Lower panel:} for the same galaxy sample, $z_\mathrm{sim}$ values are compared to estimates from standard SED fitting (\textsc{LePhare} code). Symbols are the same as in the upper panel. }
     \label{fig:zsom_cf}
 \end{figure}

 We set $N_\mathrm{c}=10$ as the scatter generated by  photometric errors typically  involves the first surrounding cells. 
 We verify that including more distant neighbours do not alter the results owing to the $1/d^2$ factor in Eq.~(\ref{eq:sfr_som}).  
 A more accurate, object-by-object determination of $N_\mathrm{c}$ could be done defining the 11-dimensional ellipsoid enclosing the 68 per cent confidence limit in all dimensions jointly. However the colours' covariance matrix is necessary for such a task  \citep{NumericalRecipes} and that is not available in \textsc{Horizon-AGN}, to be consistent with the real COSMOS catalogue.
 To tackle this limitation, \citet{hemmati19b} suggest a Monte Carlo method based on multiple realisations of the SOM mapping, extracting each time a different SED for the various galaxies (consistently with their photometric errors). A more rigorous Bayesian approach can be found in \citet{CarrascoKind14} and  \citet[][see also  \citealp{bonnett15} for neural network redshifts]{buchs19}. We postpone to future work a thorough analysis of the redshift PDF via SOM. 
 That kind of analysis shall also improve the $z_\mathrm{SOM}$ estimates by smearing out the horizontal stripes visible in Fig.~\ref{fig:zsom_cf} (upper panel). Those are the caused by a discretisation in redshift for galaxies well-segregated in a restricted area of the SOM, even though they span a large $z_\mathrm{sim}$ range. Their pseudo-observed flux in one of the filters is significantly different from the intrinsic one, despite a rather small error bar associated to it. In those cases, averaging over $N_\mathrm{c}$ neighbours is not sufficient to explore distant cells. A full Bayesian approach would possibly capture their diversity better than the present implementation.

 We emphasise that the entire procedure takes less than 30 minutes of wall clock time, whereas to process the same 371,168 galaxies \textsc{LePhare} needs more than 100 hours (without considering the computational time to estimate redshifts in the first run). 
 In Fig.~\ref{fig:zsom_cf} we provide a comparison between the true redshifts of \textsc{Horizon-AGN} galaxies and  those derived either through the SOM (upper panel) or  \textsc{LePhare}  ($z_\mathrm{phot}$, lower panel).  
 The Figure shows 371,168 COSMOS-like galaxies from $z=0$ to $\sim$3.5, namely the $S/N$-selected sample with the exception of the \textit{bona fide} galaxies used for calibration. Overall, $z_\mathrm{SOM}$ are in decent agreement with $z_\mathrm{sim}$, despite the significant scatter. 
 NMAD and outlier fraction are computed for $\Delta z\equiv z_\mathrm{SOM}-z_\mathrm{sim}$, being respectively 0.044 and 6.1 per cent.
 Galaxies at $z_\mathrm{sim}<0.2$ and  $z_\mathrm{sim}>3.2$ are the most problematic as they suffer from SOM boundary effects: at those redshifts, i.e.\ the extremes of the distribution, there are too few galaxies to train a distinct cell. For instance there are only 910 galaxies with $z_\mathrm{sim}<0.2$, spread across 23 cells; in each cell they represent  5$-$12 per cent of the objects because they are classified together with a much larger number of $0.2<z_\mathrm{sim}<0.5$ galaxies. 

The NMAD and outlier fraction we find are both larger than those computed in \citet{masters19} for the COSMOS $z_\mathrm{SOM}$, but their method slightly differs from ours as they use a deeper sample for calibration and then map (typically brighter) spectroscopic galaxies on the SOM. Here we compute the NMAD and outlier fractions with a sample that goes fainter, which can explain the slightly worse results. In Paper I we discussed the caveats of using a spectroscopically selected sub-set of galaxies to assess SED fitting quality of the parent photometric sample, as it is a limited (and sometimes biased) representation of the entire population. 
The comparison of Fig.~\ref{fig:zsom_cf} does not have this caveat because the same \textsc{Horizon-AGN} galaxies are used in both cases. We find that  \textsc{LePhare} NMAD (0.024) and outlier fraction (1.5 per cent) are significantly smaller. 
It is worth noticing that those  estimates take advantage of a more advanced Bayesian framework during the SED fitting: a galaxy $z_\mathrm{phot}$ is defined as the median of the PDF($z$) resulting from  \textsc{LePhare} template library. More complex applications of the SOM   \citep{masters19,sanchez&bernstein19} yield to redshift estimates comparable in precision to  \textsc{LePhare}.  What is nonetheless surprising  is the significantly better performance of our SOM in terms of redshift bias \citep[as pointed out also in][]{masters19} defined as the mean of $(z_\mathrm{SOM}-z_\mathrm{sim})/(1+z_\mathrm{sim})$. This can be deduced from Fig.~\ref{fig:zsom_cf} by observing the scatter in the two panels: the one of  $z_\mathrm{SOM}$ vs $z_\mathrm{sim}$ is larger but more symmetric. 
After removing outliers, the $z_\mathrm{SOM}$ bias is  -0.001, a factor 10 smaller than the bias resulting from \textsc{LePhare}.

 \begin{figure}
     \centering
     \includegraphics[width=0.99\columnwidth]{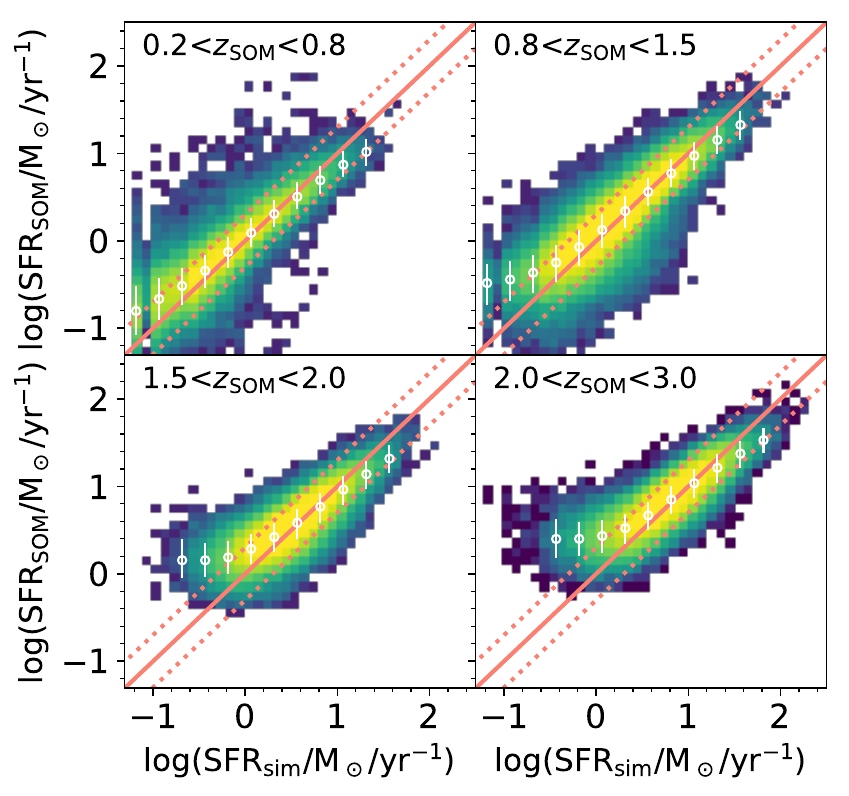}
     \caption{Comparison between intrinsic star formation rates (SFR$_\mathrm{sim}$) and estimates obtained through the SOM (SFR$_\mathrm{SOM}$) for the same \textsc{Horizon-AGN} galaxy sample shown in Fig.~\ref{fig:zsom_cf} (using also the same colour map for object density).  Each panel shows a different redshift interval: galaxies are binned according to their $z_\mathrm{SOM}$, which is provided by our method at the same time (see Sect.~\ref{subsec:measurements}).  Empty circles are median values in running bins of SFR, with error bars computed as the difference between 84th and 16th percentile. A solid line indicates the 1:1 bisector and dotted lines the $\pm0.15$\,dex offset from it.    }
     \label{fig:sfrsom_cf}
 \end{figure}
 
Along with $z_\mathrm{SOM}$, this method provides at the same time the SFR estimates.  Fig.~\ref{fig:sfrsom_cf} shows how they compare to the intrinsic SFR$_\mathrm{sim}$ in different redshift bins\footnote{
See Fig.~A4 in the On-line Supplementary Materials, for an alternate version of the calculation using $K_\mathrm{s}$ fluxes for the re-scaling in Eq.~(\ref{eq:sfr_som}).}. 
The correlation between the two is very tight for 
SFR$_\mathrm{sim}>1\,M_\odot\,\mathrm{yr}^{-1}$. Measuring its scatter as $\log(SFR_\mathrm{SOM}/SFR_\mathrm{sim})$, the NMAD is always $<$0.2 dex (i.e., smaller than the typical scatter for stellar mass estimates). Namely, in the four redshift bins of Fig.~\ref{fig:sfrsom_cf}, the logarithmic SFR scatter ranges from 0.16 to 0.18 dex. 
On the other hands, the method overestimates low levels of star formation (SFR$_\mathrm{sim}<1\,M_\odot\,\mathrm{yr}^{-1}$)  at $z>0.8$. 
There is also a systematic underestimation for the most star-forming galaxies, but that offset is always smaller than 0.15\,dex. 
The two systematics have different explanations. 
The trend  in the low-SFR$_\mathrm{sim}$ regime is due to $z_\mathrm{SOM}$ outliers:  because of SED degeneracy, a few galaxies\footnote{Note that the spatial density map used in Fig.~\ref{fig:sfrsom_cf} is in logarithmic scale.}  already in the red sequence are miss-classified in cells mostly occupied by dusty star forming galaxies at higher $z$. This bias should diminish 
in deeper surveys, as they better disentangle redshift degeneracies. 
On the opposite hand, the bending of the SFR$_\mathrm{SOM}$ vs SFR$_\mathrm{sim}$ relation for the most star forming objects is a consequence of the intrinsic SFR distribution inside those cells. Galaxies with the highest activity are in the tail of such distribution so it is unlikely that one of them is selected for calibration. The reference SFR$_\mathrm{cal}$ extracted in those cells is usually 0.1$-$0.2\,dex lower than the maximum, explaining the underestimation shown in Fig.~\ref{fig:sfrsom_cf}.

\begin{figure}
    \centering
    \includegraphics[width=0.99\columnwidth]{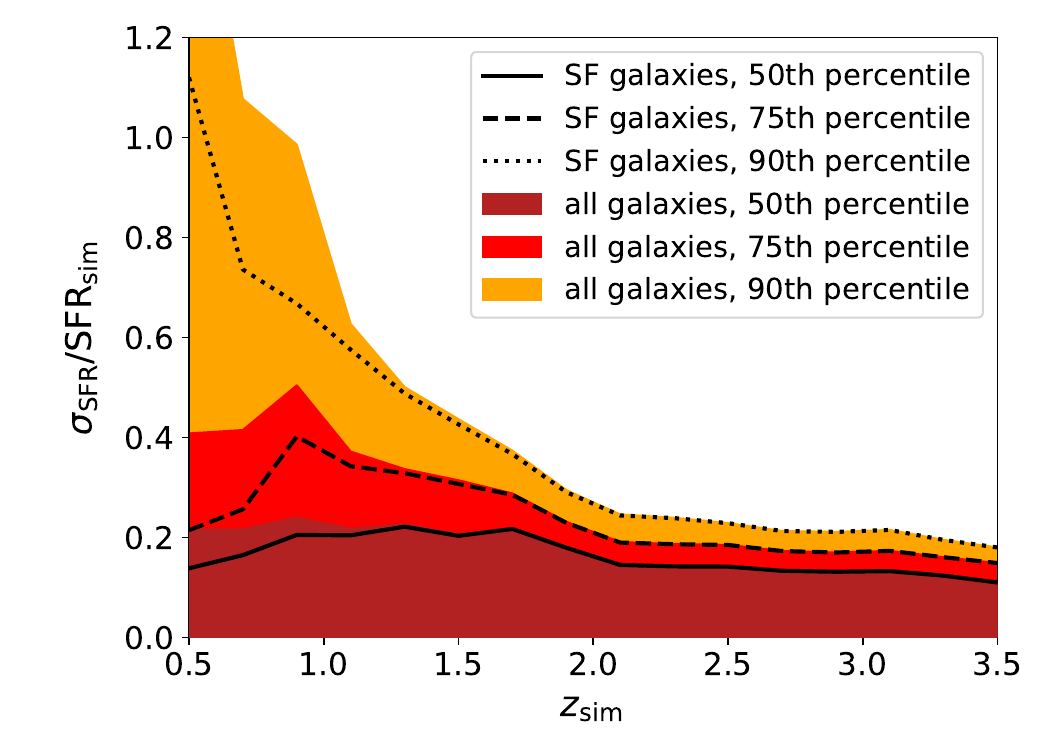}
    \caption{Fractional error of SOM-derived SFR estimates ($\sigma_\mathrm{SFR}/\mathrm{SFR}_\mathrm{sim}$) due to the SFR$_\mathrm{cal}$ stochasticity, i.e.\ the random selection of the \textit{bona fide} galaxies. To compute $\sigma_\mathrm{SFR}$ we repeat our SOM calibration procedure (Sect.~\ref{subsec:measurements}) 100 times with a different   SFR$_\mathrm{cal}$ basis (i.e., randomly re-extracting the 6,400 \textit{bona fide} galaxies) and derive  SFR$_\mathrm{SOM}$ every time. Shaded areas (from dark red to orange) show the fractional error embedding 50, 75, and 90 percent of the COSMOS-like sample as a function of redshift. Solid lines show the same quantities computed for star forming galaxies only (sSFR$_\mathrm{sim}>10^{-10}\,\mathrm{yr}^{-1}$).  
    }
    \label{fig:som_sigma_mc}
\end{figure}

We also investigate the statistical error due to the random  selection of \textit{bona fide} galaxies. The  SFR$_\mathrm{cal}$ label of a given cell might significantly change depending on which galaxy is actually targeted. A Monte Carlo simulation can quantify this uncertainty. We produce 100 calibration samples of the COSMOS-like SOM, each time randomly extracting a different set of \textit{bona fide} galaxies. The standard deviation of the SFR$_\mathrm{SOM}$ estimates ($\sigma_\mathrm{SFR}$) within the 100 realisations is calculated as a function of redshift. Remarkably, most of the galaxies at $z_\mathrm{sim}>2$ have  $\sigma_\mathrm{SFR}/\mathrm{SFR}_\mathrm{sim}<30\%$   (Fig.~\ref{fig:som_sigma_mc}). At lower redshift, $\sim$70 per cent of the sample still shows such a small uncertainty, confirming the tight SFR$_\mathrm{SOM}$-SFR$_\mathrm{sim}$ correlation discussed above. Fig.~\ref{fig:som_sigma_mc} also reveals a sub-sample of low-$z$ galaxies with a larger dispersion (although their fractional errors do not exceed a factor $\sim$2). These objects are mainly quiescent or post-starburst  galaxies in cells with a large spread in SFR   so they are more sensitive to the random   SFR$_\mathrm{cal}$ selection. 
With the same Monte Carlo we also quantify the $z_\mathrm{SOM}$ error which is always below 4 per cent.  

\section{Application of the SOM estimator to present and future surveys}
\label{sec:real_calibr}

 So far we have applied the new method without discussing the details about how to build its calibration sample in practice. 
 To collect robust measurements of their redshift and SFR, the \textit{bona fide} galaxies can be observed with present or future facilities.  We are particularly interested in the applications that our method will have in the next decade, as foreseen surveys will offer an ideal test-bed for it. This is motivated by the clear advantage of ML methods in terms of computational speed, which will be a key factor e.g.\ for future cosmology-driven missions probing large cosmic volumes. Moreover, the next generation of telescopes will allow to exploit the full potentiality of the SOM  by  assembling unprecedented calibration samples  \citep[see][]{bundy19}. In Sect.~\ref{subsec:bona_fide} we envision two of these opportunities, assuming that the SOM will be calibrated either by a large-scale spectroscopic survey in optical-NIR or from FIR observations. 
 In  Sect.~\ref{subsec:comparison} we compare the results from both calibrations to standard template fitting. The SOM method applied here is more realistic than Sect.~\ref{subsec:measurements} but is also affected by the selection function of those ``pseudo-surveys''. More details about their design and the bias they may introduce can be found in Appendix B (see the On-line Supplementary Materials).

\subsection{How to build the SFR calibration sample?}
\label{subsec:bona_fide}

 We discuss the realisation of a calibration sample within the \textsc{Horizon-AGN} framework, to be consistent with the rest of our analysis. Thanks to the wealth of observations in the COSMOS field a similar attempt can also be made using real data, although with some limitations; we postpone this test to future work. 
 We propose two alternate calibrations for the SOM, namely
 \begin{enumerate}
 \item[C1:] a spectroscopic follow-up  targeting one galaxy per cell, to derive SFR$_\mathrm{cal}$ from their H$\alpha$ flux.
 \item[C2:] a combination of UV and IR imaging covering a portion of the field, providing SFR$_\mathrm{cal}$ for several galaxies per cell via energy balance equation. 
 \end{enumerate}
 We anticipate that the analyses resulting from the two calibrations will differ, because of the specific priors  of each scenario. 
 The calibration effort also provides spectroscopic redshifts (to be used as $z_\mathrm{cal}$) but we will focus on  SFR$_\mathrm{cal}$ measurements since they introduce the major uncertainties in our SOM estimator\footnote{
 Typical spectroscopic redshift errors are sub-dominant in the present analysis therefore we assume $z_\mathrm{cal}\equiv z_\mathrm{sim}$.}. 
 Although we imagine data to be taken from next-generation facilities, galaxy parameters are assumed to be derived with the usual prescriptions. For instance, the SFR indicator adopted in  C1 follows \citet{kennicutt98}:
 
 \begin{equation}
     SFR(H\alpha) = 5.4\times10^{-42} \frac{L_{H\alpha}}{\mathrm{erg\,s}^{-1}}\; M_\odot\,\mathrm{yr}^{-1},
     \label{eq:kenni}
 \end{equation}
 in which the original coefficient ($7.9\times10^{-42}$) has been converted to Chabrier's IMF. $L_{H\alpha}$, namely the luminosity of the H$\alpha$ line, must be corrected for dust absorption. This correction can be done e.g.\ by using the Balmer decrement:  
 \begin{equation}
     E(B-V) = \frac{E(H\beta-H\alpha)}{k(H\beta)-k(H\alpha)}. 
     \label{eq:balmer_decr}
 \end{equation}
 The numerator on the right-hand side of Eq.~(\ref{eq:balmer_decr}) is the colour excess due to dust reddening \citep[see equation 2 in][]{moustakas06} while the denominator comes from an attenuation function $k(\lambda)$, as e.g.\ in \citet{cardelli89}. 
 
 With respect to the C2 case, there are different approaches in the literature to derive SFRs from UV+IR luminosity.  The one used in \citet{arnouts13} is based on the formula  
 \begin{equation}
    SFR({\rm NUV},{\rm IR}) =  8.6\times10^{-11}\frac{L_\mathrm{IR}+2.3L_\mathrm{NUV}}{\mathrm{erg\,s}^{-1}}\; M_\odot\,\mathrm{yr}^{-1},  
    \label{eq:sfr_uv+ir}
 \end{equation}
 where $L_\mathrm{IR}$ is the total IR luminosity (8$-$1000\,$\mu$m)  and $L_\mathrm{NUV}$ is the monochromatic luminosity in the near-UV rest frame filter. The IR luminosity accounts for the new-born stars enshrouded by dust that do not contribute to the NUV term. Different dust corrections have been proposed for Eq.~(\ref{eq:sfr_uv+ir}), also depending on the observations used as a proxy for $L_\mathrm{IR}$  \citep[see][]{hao11}. Further details about these SFR indicators
 can be found in  \citet[][]{kennicutt&evans12} and references therein.

  As mentioned above, we aim at designing  both calibration sample as they would be assembled by means of next-generation facilities. For instance the spectroscopic survey required for C1 could be carried out in the optical with the 4-metre Multi-Object Spectroscopic Telescope \citep[4MOST,][]{4most_msg}  and in NIR with the Multi-Object Optical and Near-infrared Spectrograph \citep[MOONS,][]{moons_spie14,moons_spie18} or 
  the Prime Focus Spectrograph \citep[PFS,][]{pfs_pasj}. In principle the \emph{James Webb} Space Telescope (JWST) could also be considered, expanding up to the mid-IR, but it is not optimised for surveying across 1\,deg$^2$ \citep[see the discussion in][]{davidzon18}. 
  We consider the case in which 4MOST and MOONS are used 
  to measure $z_\mathrm{cal}$ and SFR$\mathrm{cal}$ at $z<1.7$. For sake of simplicity we exclude higher redshifts not to rely on another nebular emission line, since this would make the calibration sample less homogeneous.
  4MOST and MOONS specifications are further discussed in Appendix B2 (On-line Supplementary Materials).

      \begin{figure*}
    \centering
    \includegraphics[width=0.45\textwidth]{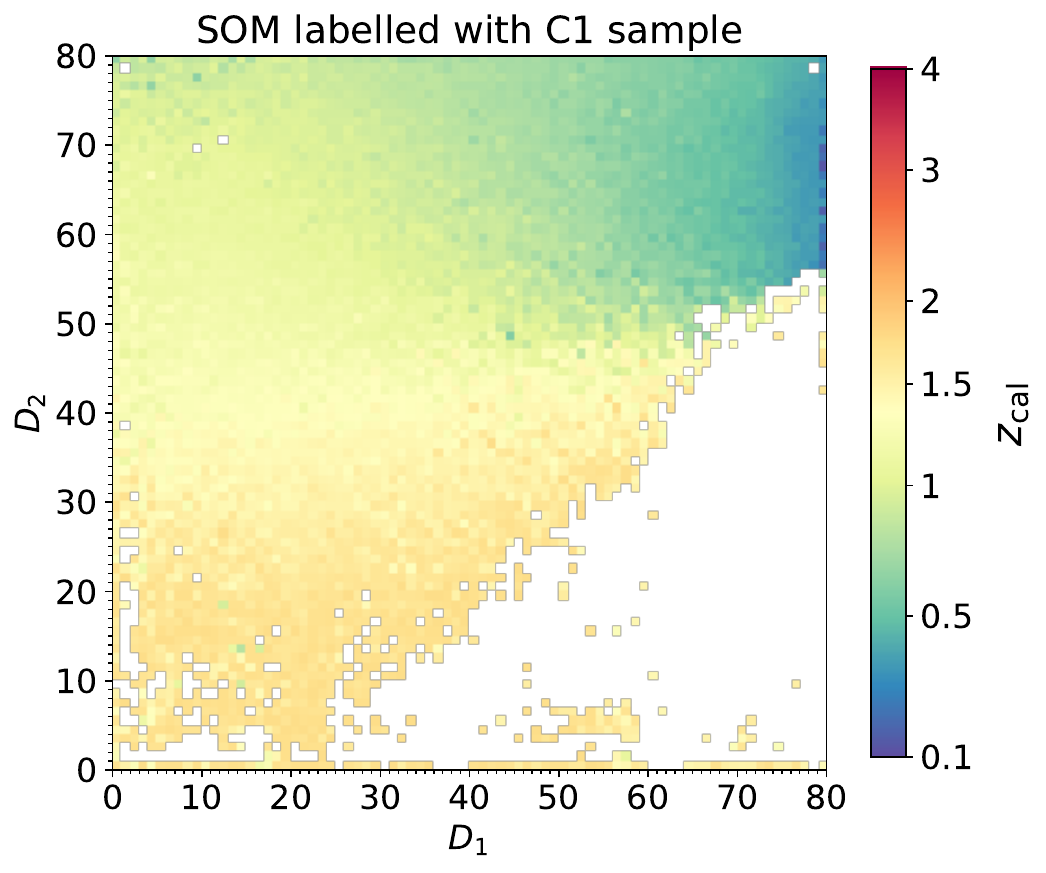}\hspace{8mm}\includegraphics[width=0.45\textwidth]{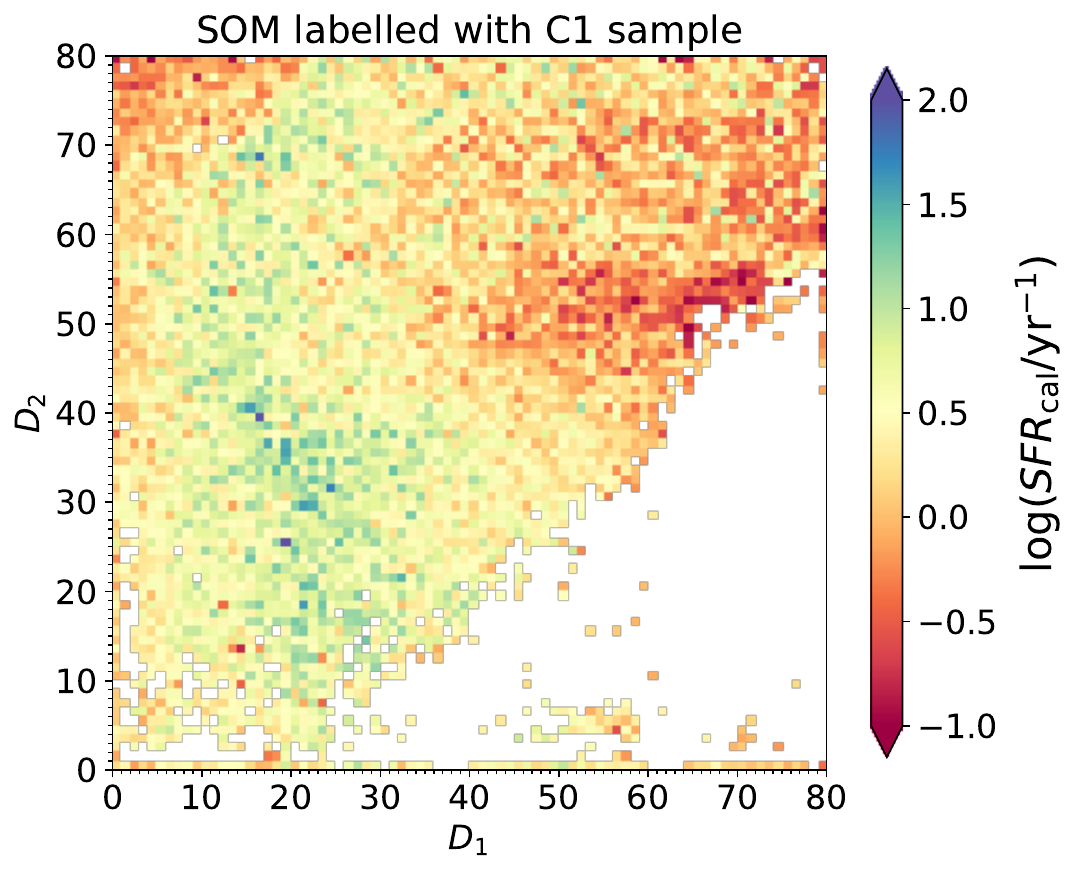}\\
   \includegraphics[width=0.45\textwidth]{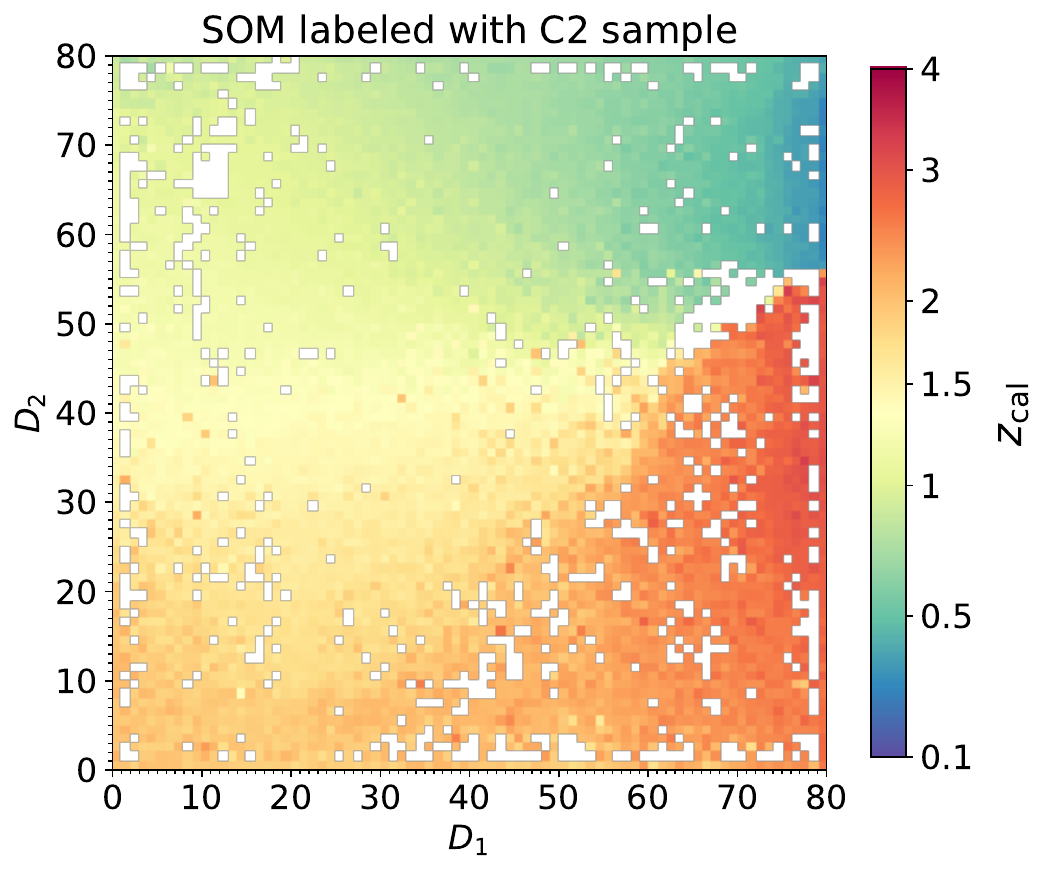}\hspace{8mm}\includegraphics[width=0.45\textwidth]{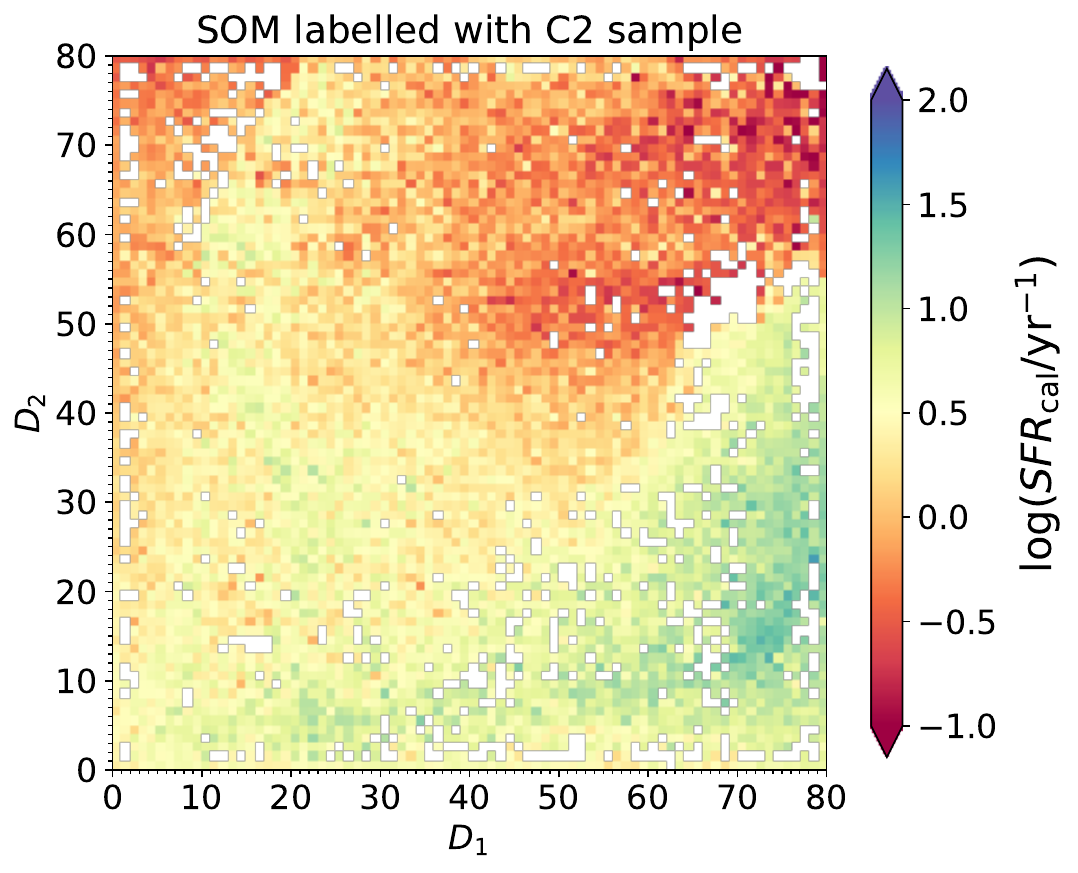}\\
    \caption{\textsc{Horizon-AGN} SOM trained with COSMOS-like colours and calibrated to work as a redshift and SFR  estimator. \textit{Upper panels:} the SOM is labelled according to the redshift ($z_\mathrm{cal}$, \textit{left panel}) and SFR (SFR$_\mathrm{cal}$, \textit{right}) of 4,749 ``spectroscopic'' galaxies coming from a pseudo-survey of calibration (C1). White pixels in the colour map are empty cells not covered by the C1 sample. \textit{Lower panels:} in this case the SOM is labelled according to $z_\mathrm{cal}$  and SFR$_\mathrm{cal}$  values (\textit{left} and \textit{right} panel, respectively) coming from an alternate calibration sample (C2). The 37,780 galaxies in this sample come from an area of 0.1\,deg$^2$ within the \textsc{Horizon-AGN} lightcone. Their SFR$_\mathrm{cal}$ is assumed to be measured from their UV and IR luminosity, that in principle could be obtained with a deep pencil-beam imaging survey.  }
    \label{fig:som_method_scheme}
\end{figure*}

 To realise the C2 sample, one could carry out  FIR observations with the proposed SPICA observatory\footnote{SPace Infrared telescope for Cosmology and Astrophysics, 
 \url{https://spica-mission.org/}} or the  Origins\footnote{https://asd.gsfc.nasa.gov/firs/} mission, both expected to launch in the 2030s. We can imagine using these telescopes to scan $\sim$0.1\,deg$^2$ of our field in the wavelength range between 20 and 230\,$\mu$m. This would result in robust $L_\mathrm{IR}$ estimates up to $z\sim3$ \citep{gruppioni17_spica,kaneda17_spica}.  To complete Eq.~(\ref{eq:sfr_uv+ir}) with rest-frame NUV luminosity one can assume to rely on GALEX data at $z<0.5$  \citep[][]{arnouts13} and deep  $u$ and $B$ photometry  at higher redshift. 
 Those data should be superseded by  higher-resolution photometry from CASTOR\footnote{CASTOR is
 the Cosmological Advanced Survey Telescope for Optical and ultraviolet Research proposed by the Canadian Space Agency \citep{castor12}. This satellite could launch as early as 2027, surveying the UV with a $\times$30 better resolution than GALEX and a $\times$100 larger field of view than HST. } and from the Large Synoptic Survey Telescope \citep[][]{lsst_sciencebook}. 
 All these future facilities are expected to observe COSMOS as one of their calibration deep fields 
 \citep{capak19b}, so in our simulated universe it is  fair to assume that a COSMOS-like lightcone can benefit from them as well.

 A difference between C1 and C2 is that the former  provides also  $z_\mathrm{cal}$ by construction, whereas  
 the C2 photometric data  must be complemented by reliable redshifts to estimate galaxy rest-frame luminosity. In the assumption of using SPICA, this shall result from its FIR high sensitivity grating spectrometer. 
 We can also suppose that the simulated  lightcone, like the real COSMOS field, will be  within the \emph{Euclid} mission footprint, so \emph{Euclid} grism redshifts will also be available\footnote{\emph{Euclid} will collect spatially resolved H$\alpha$ fluxes from  $z=0.9$ to 1.8  \citep[down to $0.5\!-\!3\times10^{-16}$\,erg\,cm$^{-2}$\,s$^{-1}$, ][]{pozzetti16} that can be used e.g.\ for aperture correction calibration of the multi-slit instruments.}. Other options are proposed in Appendix B3  along with a thorough discussion on sample variance (see On-line Supplementary Materials).    
 
To summarise, the calibration sample C1 is made by 4,749 \textit{bona fide} galaxies in an equivalent number of cells. They are supposed to be $H_\alpha$ emitters ($>2\times 10^{-17}$\,erg\,s$^{-1}$\,cm$^{-2}$) at $0<z<1.7$. C2 assumes to observe $19\arcmin\times19\arcmin$  of the \textsc{Horizon-AGN} lightcone in UV and FIR; the 40,046 galaxies in that area are stacked (binned per cell) to obtain median SFRs. 
  In the former case, the logarithmic SFR$_\mathrm{cal}$ is obtained by perturbing the original $\log(\mathrm{SFR}_\mathrm{sim}/M_\odot/\mathrm{yr}^{-1})$ of each \textit{bona fide} galaxy with random Gaussian noise.  
  The Gaussian standard deviation  is set to  $\sigma=0.18$\,dex from comparison to state-of-the-art surveys  \citep[e.g.\ FMOS-COSMOS,][]{kashino19}. 
  In C2 we do not attempt to reconstruct  $L_\mathrm{UV}$ and $L_\mathrm{IR}$ for sake of simplicity. 
  The SFR$_\mathrm{cal}$ of a given cell is the median $\mathrm{SFR}_\mathrm{sim}$ of the \textit{bona fide} galaxies inside it\footnote{
  This is expected to be a good proxy of the UV+IR  estimator, whose timescale is similar to the 100\,Myr interval used to define SFR$_\mathrm{sim}$ in \textsc{Horizon-AGN} \citep[it is also comparable with H$\alpha$-derived measurements, see][]{kashino19}.}, perturbed with Gaussian noise \citep[$\sigma=0.1$\,dex, see][]{ilbert15}. 
  Eventually, the SOM is labelled with the $z_\mathrm{cal}$ and  SFR$_\mathrm{cal}$ values of either 
  C1 (Fig.~\ref{fig:som_method_scheme}, upper panels) or C2 (lower panels).  Depending on the used \textit{bona fide} sample, certain cells do not get a label.

 \begin{figure*}
     \centering
     \includegraphics[width=0.9\textwidth]{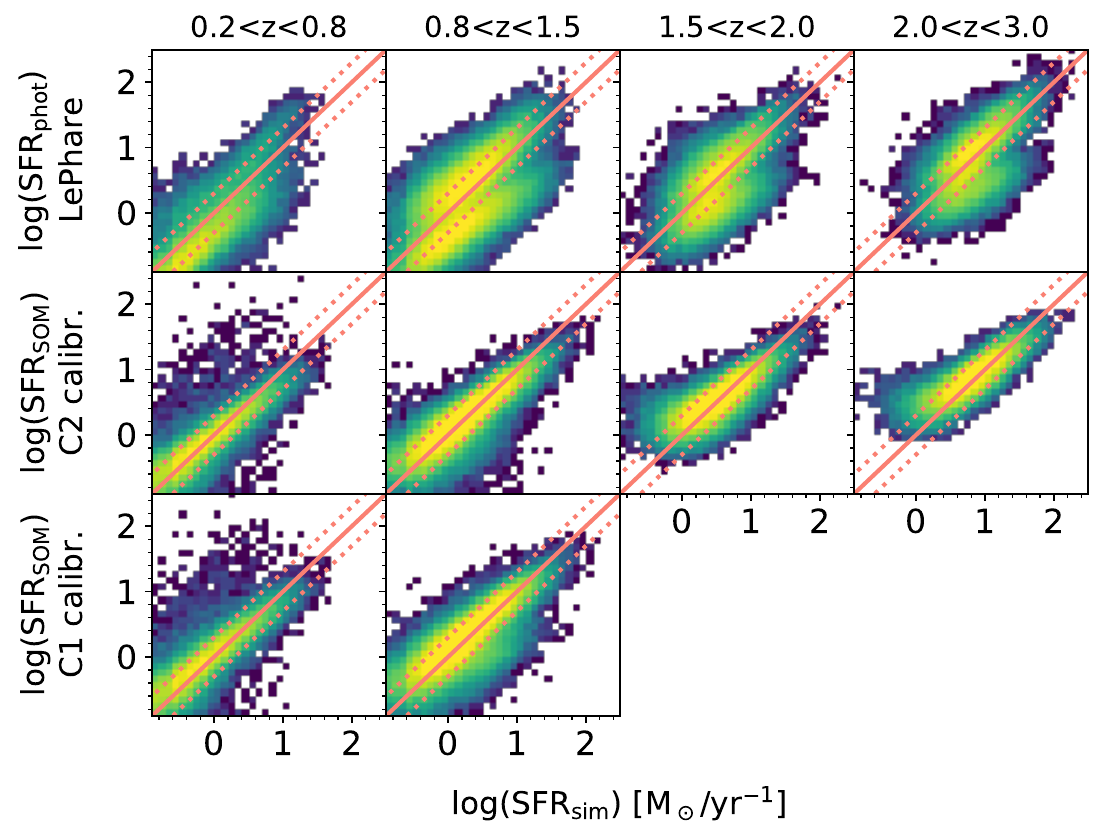}
     \caption{Comparison between intrinsic SFR$_\mathrm{sim}$ vs estimates obtained in from different methods: standard template fitting using the code \textsc{LePhare} (\textit{upper row of panels}); using the SOM method with the calibration described as version C2 in the text (\textit{middle panels}); SOM method with a different calibration sample, referred as C1 in the text (\textit{lower panels}). Galaxies are binned in the different redshift bins according to either their $z_\mathrm{phot}$ from  \textsc{LePhare} or $z_\mathrm{SOM}$ in the case of the new method proposed here.  Version C1 is limited to the first two $z$ bins because of the way the calibration sample is constructed. In each panel the solid line indicates the 1:1 relationship and the dotted lines a $\pm0.15$\,dex offset from it. }
     \label{fig:sfrsom_cf1}
 \end{figure*}

\subsection{SFR estimates and comparison with template fitting} 
\label{subsec:comparison}

 After labelling  the SOM  we apply the procedure described in Sect.~\ref{subsec:measurements} to assign a SFR$_\mathrm{SOM}$ estimate (Eq.~\ref{eq:sfr_som}) to each photometric galaxy.   
 The outcome can be compared to that obtained in Paper I by using \textsc{LePhare}. We do not show the $z_\mathrm{SOM}$ vs $z_\mathrm{sim}$ comparison as the trend is similar to Fig.~\ref{fig:zsom_cf} (upper panel). Despite the additional uncertainties introduced in Sect.~\ref{subsec:bona_fide}, the figure of merit does not change. For the C2 calibration, which covers the same redshift range of Fig.~\ref{fig:zsom_cf},  NMAD$(z_\mathrm{SOM})$ and outlier fraction remain 0.043 and 6 per cent. In the following we focus on the   SFR$_\mathrm{SOM}$ results, which show a remarkable improvement with respect to template fitting estimates (SFR$_\mathrm{phot}$). 
 
 We remind that the estimate of physical properties via  template fitting involves a two-step procedure. First, to find their $z_\mathrm{phot}$, \textsc{LePhare} fits galaxy SEDs with a composite set of templates \citep[described in][]{laigle16}; then, after fixing the redshift of each galaxy to $z_\mathrm{phot}$, the code calculates the SFR (along with stellar mass and other physical quantities) by means of another SED library, this time made from BC03 models. Since our mock catalogue reproduces COSMOS2015,   \textsc{LePhare} is used with the same configuration as in \cite{laigle16} . More details about running \textsc{LePhare} to estimate SFRs in \textsc{Horizon-AGN} can be found in Paper I.

 \begin{figure}
     \centering
     \includegraphics[width=0.99\columnwidth]{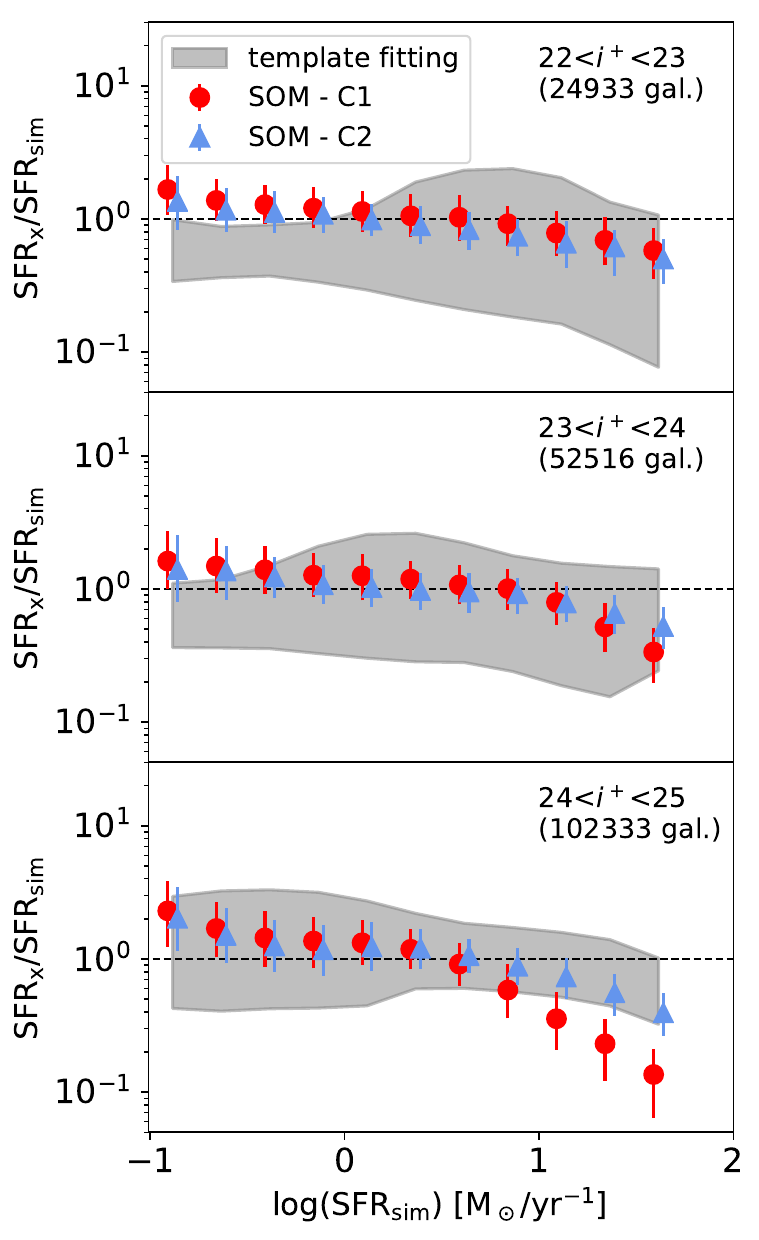}
     \caption{Median offset between intrinsic SFR and SOM-based estimates, for \textsc{Horizon-AGN} galaxies with COSMOS-like photometry (each panel showing a different $i^+$-band magnitude range). The SFR$_\mathrm{SOM}$ estimates are derived wither using the SOM calibration C1 (red circles) or C2 (blue triangles); symbols are horizontally shifted of $-0.03$ and $+0.03$\,dex respectively  for sake of clarity. Error bars delimit the 16th-84th percentile range of each SFR$_\mathrm{SOM}$/SFR$_\mathrm{sim}$ distribution. Estimates from \textsc{LePhare} are also included, but given the bimodality of   SFR$_\mathrm{phot}$/SFR$_\mathrm{sim}$ (see Fig.~\ref{fig:sfrsom_cf}, upper panels) The figure shows only the intervale between 16th and 84th percentile in running bins of SFR$_\mathrm{sim}$.  In each panel a dotted line marks the 1:1 ratio. }
     \label{fig:sfrsom_cf2}
 \end{figure}
 
 The upper panels of Fig.~\ref{fig:sfrsom_cf1} shows the comparison between  SFR$_\mathrm{phot}$ and  SFR$_\mathrm{sim}$ in different bins from $z_\mathrm{phot}=0.2$ to 3. Under- and over-estimates produced by \textsc{LePhare} are clearly visible, generating a two parallel sequences  in the distribution. Such a bimodality is due to SED fitting degenarcies. By comparing a dust-free ve dusty  Universe, Paper~I isolated the major role of dust attenuation in driving this bimodality. In particular, the choice of the extinction curves in the template library is pivotal. This remains true even when the redshift is fixed to its intrinsic value instead of $z_\mathrm{phot}$. Inadequate extinction models or $E(B-V)$ values may cause indeed an overestimation or an underestimation of the SFR, as shown in Appendix~B of Paper~I.  

 The same photometric galaxies\footnote{
 All the \textit{bona fide} galaxies used for calibration have been excluded from the comparison.} are reported in the middle and lower panels of Fig.~\ref{fig:sfrsom_cf1} comparing intrinsic SFRs with the values computed through the SOM (C2 and C1 version respectively). The performance of the SOM is significantly better than template fitting. The SFR$_\mathrm{SOM}$ distribution does not show the same bimodality observed for the  SFR$_\mathrm{phot}$ estimates because the SOM  fitting is based only on \textit{observed} SEDs, which naturally include the ``correct'' dust attenuation law and $E(B-V)$ range. 
 In \textsc{LePhare} the grid of templates is built without strong  observational priors so the library is affected by artificial degeneracies. 
 
 Also our method is model dependent because of the SFR$_\mathrm{cal}$ labels. These measurement requires some theoretical prescription (e.g., about dust attenuation or IMF). However we argue that the required assumptions, for either  C1 or C2, introduce a milder bias than template fitting. For instance the C1 calibration requires the choice of a dust extinction curve  (Eq.~\ref{eq:balmer_decr}) but the difference between models is small: e.g., $k(H\beta)-k(H\alpha)=1.07, 1.27$ for \citet{cardelli89} and  \citet{calzetti2000}  respectively.  
  On the contrary \textsc{LePhare} templates are constrained by data at bluer r.f.\ wavelengths where the dust extinction curve plays a more important role \citep{ilbert09}; for the same models in the example above: $k(2000\textup{\AA})-k(3000\textup{\AA})=3.18$ and $1.95$. 
  
 With respect to the C2 sample one may notice that the energy balance equation is also implemented in template fitting codes \citep[e.g.\ \textsc{MAGPHYS},][]{dacunha08} sometimes in very elaborated ways including also the AGN contribution \citep[e.g.\ SED3FIT,][]{berta13a}. In fact one may also use one of those codes  instead of Eq.~\ref{eq:sfr_uv+ir} but not for the other  \textsc{Horizon-AGN} galaxies that do not belong to C2.
 Moreover, even if the whole galaxy sample were observed in UV and FIR, codes like \textsc{MAGPHYS} are extremly expensive in terms of computational time and run only after fixing the redshift. This 2-step fitting procedure, which is of widespread use  in the literature, raises several issue \citep[e.g., propagation of $z$ uncertainties,][]{grazian15,davidzon17}. On the other hand, the SOM does not require such a procedure, providing $z$ and SFR estimates simultaneously.  

 Fig.~\ref{fig:sfrsom_cf2} proposes the same comparison of Fig.~\ref{fig:sfrsom_cf1} in a different flavour, i.e.\ showing the median ratio  SFR$_\mathrm{SOM}$/SFR$_\mathrm{sim}$ in three bins of apparent magnitude from $i^+=22$ to 25. 
 At $i^+<23$ (Fig.~\ref{fig:sfrsom_cf2}, middle panel) there is an excellent agreement of both C1 and C2 estimates with the intrinsic SFR. Such a trend is still observed at $23<i^+<24$ even though the most star forming galaxies start to be systematically underestimated by more than a factor 2.  This is a border effect inherent to the SOM analysis already discussed in Sect.~\ref{subsec:measurements}; having a larger sample that allows for a more representative SOM would mitigate this effect \citep{buchs19}.  The discrepancy becomes  more accentuated at $24<i^+<25$ (Fig.~\ref{fig:sfrsom_cf2}, lower panel) especially for the C1 calibration that  by construction relies on \textit{bona fide} galaxies systematically brighter than the average   \citep[an effect that could  be accounted for by a higher sampling rate in those cells, see][]{masters19}. 
 Concerning \textsc{LePhare}, under- and over-estimates would compensate each other resulting in a misleading  SFR$_\mathrm{phot}$/SFR$_\mathrm{sim}\simeq1$. Therefore 
 Fig.~\ref{fig:sfrsom_cf2} does not show the median of the SFR$_\mathrm{phot}$ distribution but only the interval between the 16th and 84th percentile. Such a dispersion is significantly larger than the ML estimates in all the magnitude bins.


\section{Summary and conclusion}
\label{sec:conclusion}

 Compared to the large number of studies measuring galaxy redshifts with  ML techniques, little progress has been made concerning other physical parameters. In spite of that, ML methods will be pivotal in the near future to derive stellar mass, star formation history, and other galaxy properties in extremely large data sets from surveys such as \emph{Euclid} and LSST. 
 In addition to their unprecedented
 speed, these algorithms (particularly the unsupervised ML methods) may lead to a ``new paradigm'' in which human intervention (i.e., the application of interpretative models) starts after galaxy classification and demographics have been decided by the machine. 
 However, results may be affected by new kinds of systematics   introduced, e.g.,  during the data reduction process or the  training set selection. A thorough investigation of ML performance and the role of its ``observational priors'' is thus imperative before such high expectations may be deemed justified.

With this in mind, we have explored advantages and limitations of the SOM as a galaxy parameter estimator  independent of model templates. We chose the SOM  because it is  an unsupervised  dimensionality reduction algorithm able not only to learn the complex structure of data but also to project it in a lower-dimensional space (2-d in our analysis) still preserving its ``topological'' features. 
 It should be clarified that our goal is not advocating for the SOM to replace standard template fitting: the two \textit{complementary} approaches should be used in synergy, the same way semi-analytic and hydrodynamical simulations have contributed to inform each other and together improve our understanding of galaxy evolution. For example, ML investigations may help to shed light on the systematic $M_\ast$ underestimation in unrisolved SED fitting of star froming galaxies 
 \citep{pozzetti07,sorba&sawicki15,sorba&sawicki18}. 
 
 We tested the SOM with a mock galaxy catalogue \citep[presented for the first time in][Paper I]{laigle19} derived from the \textsc{Horizon-AGN} hydrodynamical simulation. Galaxies cover 1\,deg$^2$ area and a redshift range $0<z<4$, with a photometric baseline similar to state-of-the-art surveys (broad-band filters from $u$ to [4.5$\mu$m]). The SOM has been trained using as input only galaxy colours, to be an analogue of 
 ``classical'' SED fitting codes like \textsc{LePhare} or \textsc{EAZY} \citep{brammer09}. In principle other (e.g., morphological) features may be trained for, but this is left to future work. 
 After classifying the mock galaxies in about 6,400 different classes (called ``cells'' in our jargon) we 
 explored the connections between the class/cell a galaxy belongs to and its physical properties. Then, we calibrated \textit{a posteriori} the SOM by labelling each cell with a value of $z$ and SFR, so that other galaxies in the same cell may have a proxy of their own redshift and star formation activity. Eventually we used the calibrated SOM to estimate the SFR for a sub-sample of about 375,000 mock galaxies between $z\sim0$ and $3$. 
 Our findings are summarised in the following.

\begin{itemize}
\setlength\itemsep{0.5em}

    \item  The SOM is an effective tool to visualise the characteristics of a complex, \textit{non-linear} 
    manifold as the \textsc{Horizon-AGN} lightcone. Galaxies are organised in a (human-readable) 2-d grid without smearing out the features of their original parameter space. 
    Moreover this is computational inexpensive, suggesting a convenient way to describe and inspect the properties of extremely large simulations \citep{mitra15}.     
    
    \item Since in our case the parameter space is made by  observer's frame colours, the SOM works like an SED fitting algorithm without a pre-compiled library of templates: the SOM  adapts its cells/weights to the data so that galaxies with similar colours (i.e., similar SEDs) are enclosed the same cell. We find that also the high-resolution spectra turn out to be nicely classified, in spite of using only broad-band photometry for training. 
    
    \item  We confirm that objects in the same cell have similar redshift (as shown in \citealp{masters15} in the observed universe) but we also find that their $M/L$ and sSFR is similar, with typical scatter of 0.15 and 0.3\,dex respectively. 
    Also  $M$ and SFR, after taking into account a normalisation factor, are well correlated to the cell where a galaxy lies. After including photometric uncertainties (modelled after the  COSMOS2015 survey) and rejecting objects with $S/N<1.5$ in any band, we trained again the SOM: the correlation between galaxy properties and cells was still present, although with larger scatter. This indicates that our analysis can be reproduced in real optical-NIR surveys (provided a sufficient depth of the observations). 
    
    \item We have measured the redshift of COSMOS-like ($S/N>1.5$) galaxies  through the SOM, finding a fairly good agreement with intrinsic $z_\mathrm{sim}$ but a larger scatter than template fitting: The $z_\mathrm{SOM}$ vs $z_\mathrm{sim}$ comparison results into NMAD$\,=0.044$ and about 6 per cent of catastrophic errors, whereas with \textsc{LePhare} they are respectively 0.024 and 1.5 per cent. On the other hand the redshift bias in the SOM case is significantly smaller  ($-0.001$, compared to $-0.011$ in \textsc{LePhare}). We considered such a result sufficiently good for our purposes so we did not attempt to improve the redshift estimator (as done e.g.\ in \citealp{buchs19}).  
    
    \item Exploiting these SOM properties, we have developed a new SFR estimator for photometric galaxies. We have assumed that a small fraction of them (10 per cent or less) has been followed-up to serve as a calibration sub-sample, providing labels ($z_\mathrm{cal}$, SFR$_\mathrm{cal}$) to the SOM cells. We have discussed possible follow-up strategies with optical-NIR spectroscopy or with UV+FIR instruments, and the possible bias introduced by each of them. After accounting for such uncertainties we have compared the  SFR$_\mathrm{SOM}$ of COSMOS-like galaxies with their intrinsic SFR$_\mathrm{sim}$. Overall the dispersion (defined as the range between 16th and 84th percentile in logarithmic bins of SFR) is  $\pm0.2$\,dex, with a small systematic offset (median $\log(\mathrm{SFR}_\mathrm{SOM}$/SFR$_\mathrm{sim})\simeq0.02\!-\!0.04$\,dex). The most active galaxies are an exception, being significantly underestimated  because they are in cells whose majority of objects (including the calibration ones) are less star forming. 
    
    \item \textsc{LePhare} SFR$_\mathrm{phot}$ estimates are also available in \textsc{Horizon-AGN} and we have compared them to the new indicator. The latter performs remarkably better: SFR$_\mathrm{SOM}$ are more precise but also significantly less biased, as they do not rely on a template library that introduces \textit{artificial degeneracies} in the SED fitting (as investigated in Paper I). 
        
\end{itemize}

    The suboptimal performance of the SOM as a redshift machine found in this analysis is partly due to the fact that we have not entirely followed \citet{masters15} prescriptions, e.g.\ we did not use cell occupation as a prior nor we distinguished a deeper calibration sample from the rest of the survey (or use a combination of multiple fields). We note that the comparison is not straightforward since  \citeauthor{masters15}, working with observed galaxies, are forced to use a spectroscopic sub-sample that is biased to some extent (see Paper I). However, as highlighted in \citeauthor{masters19}, on that spectroscopic sample their $z_\mathrm{SOM}$ figure of merit is better than \textsc{LePhare}.

    On the other hand, the better performance of our SOM method to compute SFR does not imply that it is bias-free: some systematics may be introduced while selecting the calibration galaxies and measuring their SFR$_\mathrm{cal}$. We argue however that model assumptions  in the SFR$_\mathrm{cal}$ calculation are generally less severe than those involved in the construction of libraries from stellar population synthesis models, with a coarse grid of $E(B-V)$ values, simplistic SFHs, fixed stellar metallicity, and other limitations to which SFR is sensitive \citep[][see also discussion in Paper I]{papovich01}. 
    Moreover if the sub-sample used for calibration turns out to be strongly biased it can be replaced by a better one without re-classifying the target galaxies, while any improvement in the  template library of \textsc{LePhare} would require to run again that (computationally expensive)  code over the whole catalogue. It should also be emphasised that  estimates of redshift and physical parameters are provided simultaneously -- a unique advantage of the SOM method that in future developments shall allow for a better treatment of $z$ error propagation. 
    
    We aim at transferring our method to the real COSMOS catalogue in the next paper of this series, even though data available in that field may be able to calibrate
    the SFR only in a limited portion of the SOM. Nonetheless, this effort can result in an original comparison between different estimators. For example, one could  derive SFR from radio continuum stacking \citep[as in][]{karim11} vs UV+IR luminosity \citep[as in][]{ilbert15} for galaxies in the same cells, easily identifying the region of the parameter space where the indicators disagree. 
    
    The present work is also intended to provide a new science case for upcoming large-area surveys. The SOM requires an accurate calibration sample 
     relatively modest in size \citep[but large enough to limit  sample variance effects,][]{buchs19}) and then billions of galaxies (e.g.\ from the 15,000\,deg$^2$ of \emph{Euclid}) can be efficiently mapped to get a proxy for their redshift \textit{and} physical properties. This is particularly true for the \emph{Euclid} Deep Fields, which will have a photometric baseline similar to the one assumed here thanks to the complementary surveys in optical (Hawaii two-O, PI: D.~Sanders) and MIR (\emph{Euclid}/WFIRST \emph{Spitzer} Legacy Survey, PI: P.~Capak). 
    We also mentioned the contribution that 4MOST, MOONS, and PFS may provide to calibrating the galaxy colour space, owing to their unprecedented multiplexing. Our case study also supports the concept of a deep surveying of COSMOS with CASTOR, SPICA, and Origins, to continue its use as a reference field for the coming decades.

\section*{Acknowledgements}
{\sl 
 ID thanks Stefano Andreon, Sirio Belli, Micol Bolzonella, Keerthana Jegatheesan, Chris Hayward, Lucia Pozzetti for useful discussions, and Elvira 
 CL is supported by a Beecroft Fellowship. 
 OI acknowledges the funding of the French Agence Nationale de la Recherche for the project ``SAGACE''. 
 This research was supported in part by the National Science Foundation under Grant No. NSF PHY-1748958 and by the NASA ROSES grant 12-EUCLID12-0004. 
 The analysis presented in this work  relied on the HPC resources of CINES (Jade) under the allocation 2013047012 and c2014047012 made by GENCI and on the Horizon and CANDIDE clusters hosted by Institut d'Astrophysique de Paris. We warmly thank S.~Rouberol for maintaining  these clusters on which the simulation was post-processed. 
 This research is part of  ERC grant 670193 and 
 {\sc horizon-UK} and is also partly supported by the Centre National d'Etudes Spatiales (CNES). 
 Several \texttt{python} packages were used, including \textsc{ASTROPY} \citep{astropy2013,astropy2018} and \textsc{SOMPY}   (main  contributors: Vahid Moosavi, Sebastian Packmann, Iv\'an V\'allas). 
}
\bibliographystyle{mnras}
\bibliography{papers}

\appendix

\section{Caveats in the SOM of \textsc{Horizon-AGN} galaxies}
\label{appendix1}

 \subsection{Resolution and subgrid recipes in {\sc Horizon-AGN}}\label{appendix1_theo}
 As mentioned in Section~3.3, our modelling of  {\sc Horizon-AGN}  photometry is limited both by the resolution of the simulation and the accuracy of the recipes implemented at the subgrid scale. 
 The spatial resolution of {\sc Horizon-AGN} simulation is at best 1~pkpc, and the mass resolution  is  $\sim 8\times10^7\,{\rm M}_\odot$ for dark matter particles. In \citet{dubois14} this translates into a lower limit  of 
 $10^8\,{\rm M}_\odot$ in galaxy stellar mass,  while  we stop at $M=10^9\,{\rm M}_\odot$ to be conservative. Our galaxies are therefore resolved by at least $\sim 500$~particles. However, in despite of the SFH of each galaxy being well sampled, the scale limit hinder us from accounting for radiative transfer and for the impact of the turbulence on star-formation. It  will naturally smooth out clumpiness in the ISM and therefore in the SFH themselves. One can expect that intrinsic SFH and metallicity enrichment histories of our simulated galaxies are less diverse at low than high resolution \citep[see e.g. ][for a study on the limited effect of resolution on the gas mass fraction]{lagos16}. 
 
  \begin{figure*}
     \centering
     \includegraphics[width=0.95\textwidth]{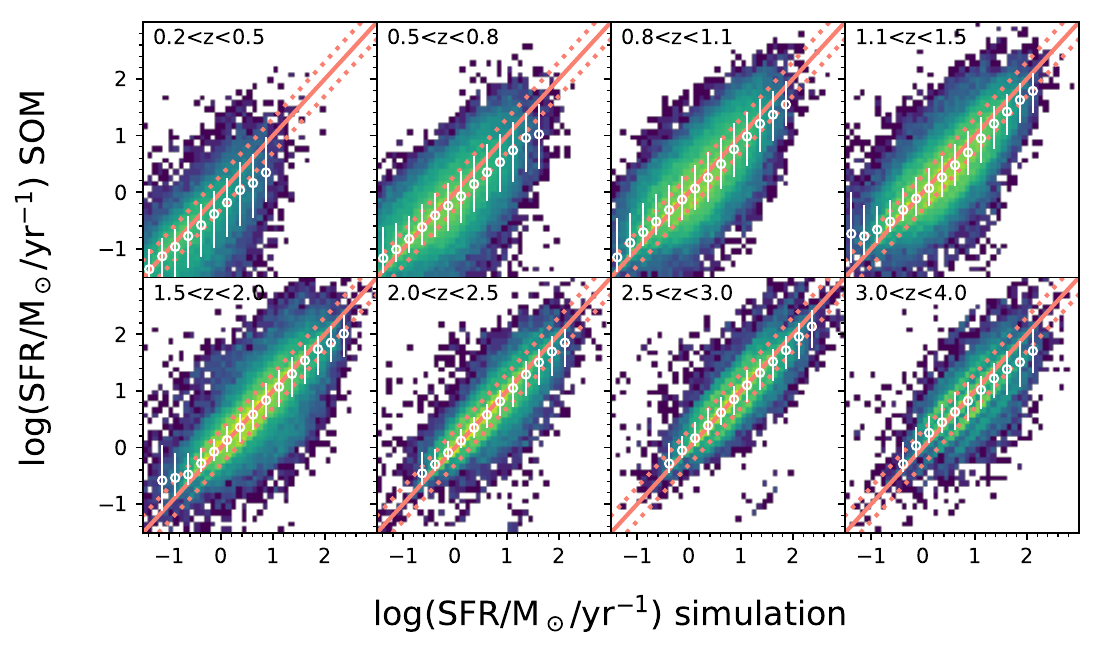}
     \caption{Comparison between the intrinsic SFR and the estimates derived from the SOM for 400,000 mock galaxies with $10^8<M/M_\odot<10^{12}$ and $0<z<4$. These objects are selected from an empirical simulation that reproduces the observed COSMOS2015 galaxies by means of their best-fit templates.  Each panel in the figure show the comparison in a different redshift bin, as indicated. A solid line marks the 1:1 bisector while dotted lines are offset by $\pm0.15$\,dex from it. Empty circles show the median in running bins of SFR$_\mathrm{sim}$ with error bars derived from the 16th$-$84th percentile range. }
     \label{fig:sfr_empirical_simu}
 \end{figure*}

 \subsection{Modelling of the photometry}\label{appendix1_photo}
 Beyond these intrinsic limitations of the simulation, we made also several simplifying assumptions while computing galaxy photometry in post-processing. These shortcomings are detailed in Appendix~A.5 of Paper~I and listed here for completeness. We assume a single and spatially constant IMF and stellar mass-loss prescription as in BC03. Dust distribution is assumed to follow the gas metal distribution, and we take a dust-to-metal ratio constant in time and space. For all these reasons, the simulated photometry naturally presents less variety than the observed one, and the SOM performance that we derive must be considered as optimal estimates. 
 
 Besides that, nebular emission is not taken into account when mimicking the COSMOS photometry. Emission lines within the wavelength range of a broad-band filter can boost the measured flux and alter galaxy colours. However, nebular emission should not impair the SOM estimator 
 as flux contamination is proportional to galaxy star formation activity. The resulting SOM would have a different appearance because of the modified colours, but without loosing the capacity to classify galaxy SEDs.  
 Both the $10^9\,{\rm M}_\odot$ lower limit and the lack of emission lines are tested by means of an empirical simulation built independently of \textsc{Horizon-AGN}. In this simulation we fit BC03 models to COSMOS2015 galaxies, fixing the redshift to the $z_\mathrm{phot}$ value provided there. We integrate each best-fit BC03 spectrum within the same filters used in the present work, adding realistic noise to get a photometric catalogue similar to the observed one. The BC03 spectra include UV-optical emission lines  (e.g., Ly$\alpha$, [OII]$\lambda$3727, H$\beta$, [OIII]$\lambda\lambda$4959,5007,
 H$\alpha$) according to \citet{schaerer&debarros09} recipe. Owing to the deep IR data, the COSMOS2015 catalogue includes numerous objects with $M<10^9\,{\rm M}_\odot$ \citep[see fig.~17 in][]{laigle16} and so does this replica.
 We produce the SOM of the empirical simulation using the empirically simulated galaxies between $M=10^8$ and $10^{12}\,{\rm M}_\odot$, then we apply our method to calculate their SFR. The resulting values are in good agreement with the intrinsic SFR (Fig.~\ref{fig:sfr_empirical_simu}). Despite the simplistic  approach (e.g., BC03 models assume $tau$ or delayed-$tau$ SFHs) this test indicates that the SOM estimator should work well also for a survey encompassing a larger stellar mass range and whose galaxy photometry is contaminated by nebular emission.  
 
\begin{figure}
    \centering
    \includegraphics[width=0.99\columnwidth]{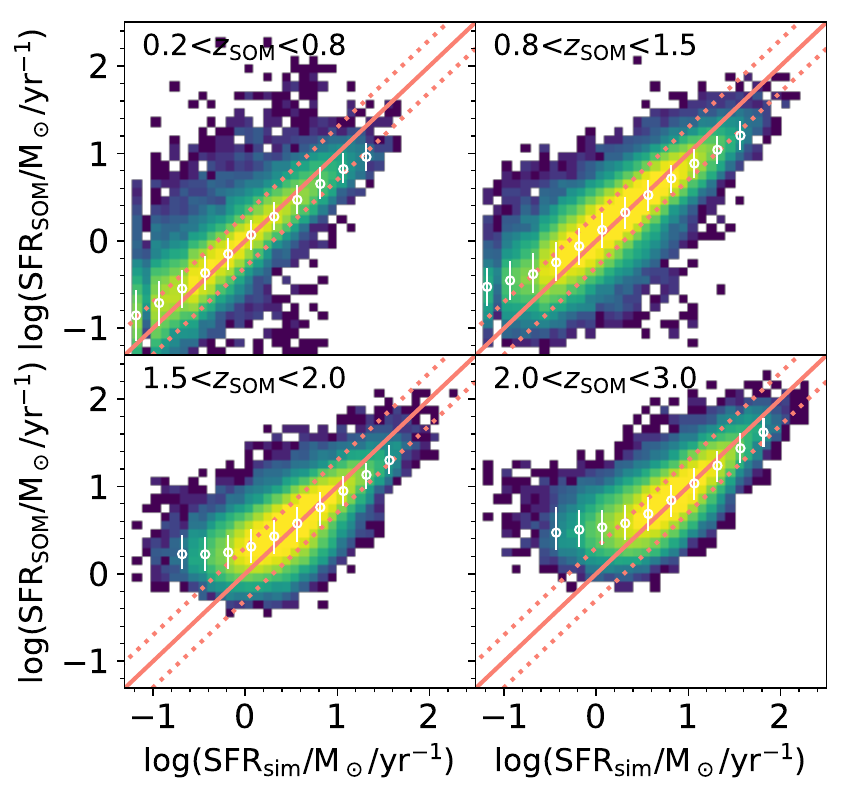}
    \caption{SFR$_\mathrm{SOM}$ vs SFR$_\mathrm{sim}$ comparison, similar to Fig.~9 (see main text) but normalising the SEDs as a function of their $K_\mathrm{s}$ flux instead of $i^+$.  }
    \label{fig:sfrsom_cf_ks}
\end{figure}

 We also test the $i^+$ flux normalisation used in the computation of stellar mass and SFR from the SOM (e.g., Fig.~5 in the main text). Since galaxy SEDs in the same cell have similar shape, the normalisation factor (see Eq.~4 in the main text) does not change significantly if another band is used (with the exception of the bluest filters). Fig.~\ref{fig:sfrsom_cf_ks} shows this 
 is the case for a $K_\mathrm{s}$ flux normalisation. In the main text we rely on the $i^+$ because it is has a higher $S/N$ whereas photometric errors in $K_\mathrm{s}$ would produce additional scatter in the SFR$_\mathrm{SOM}$ vs SFR$_\mathrm{sim}$ comparison (cf.\ Fig.~\ref{fig:sfrsom_cf_ks} with Fig.~9 in the main text). 
 
\section{Training and calibration galaxy samples}

 \subsection{Training sample selection}
 
  \begin{figure}
     \centering
     \includegraphics[width=\columnwidth]{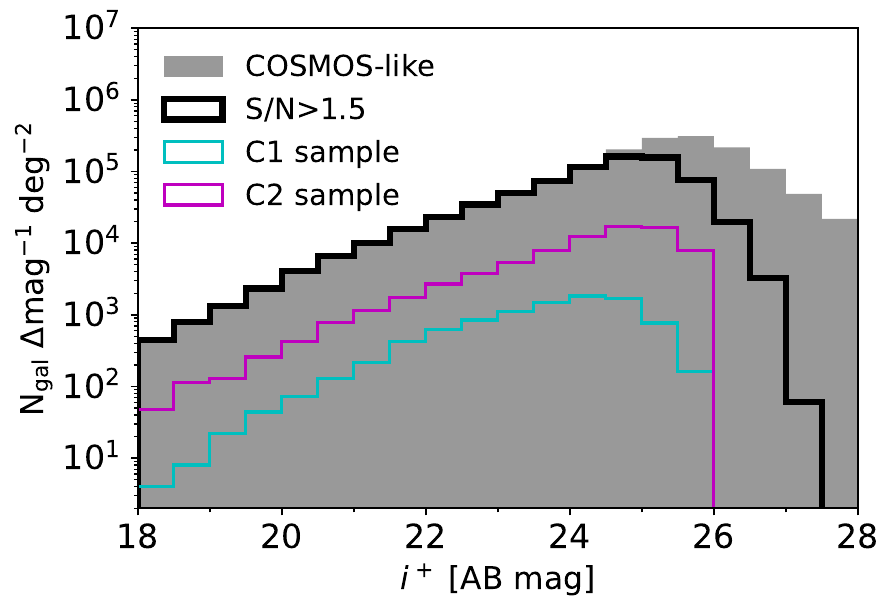}\\
     \includegraphics[width=\columnwidth]{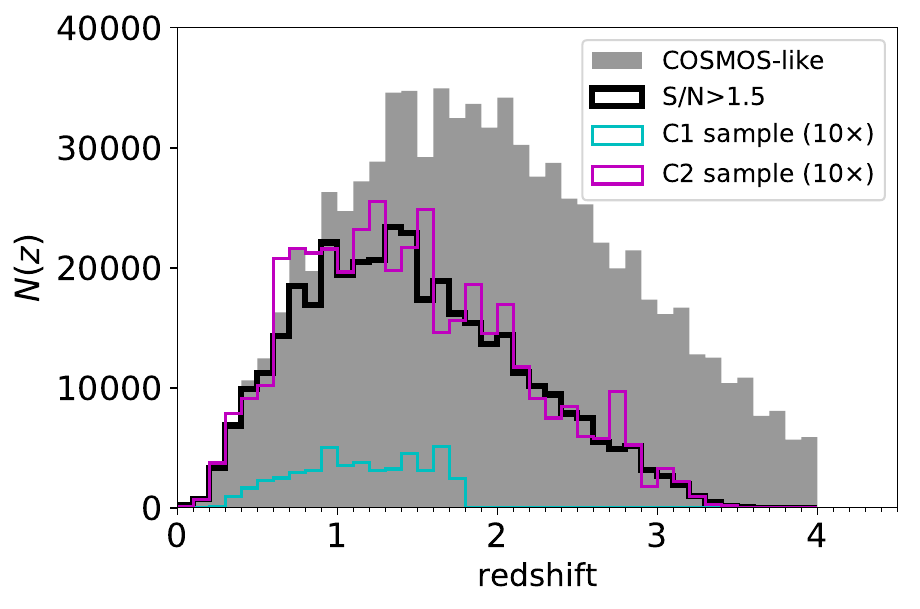}
     \caption{Magnitude and redshift distribution (\textit{upper} and \textit{lower} panel respectively)    of \textsc{Horizon-AGN} galaxies with COSMOS-like uncertainties in their photometry. 
     Both panels show the distribution of the  parent  sample (filled histogram), the $S/N$-selected sample (black line, see Sect.~4.1), and the objects extracted to label the SOM in the C1 (cyan) and C2 (magenta) scenarios (see Sect.~5.1 for more detail about the two pseudo-surveys). In the bottom panel, number counts of C1 and C2 sub-samples are boosted by a factor 10 to improve readability. 
     }
     \label{fig:som_method_hist}
 \end{figure}
 
\begin{figure}
     \centering
     \includegraphics[width=0.99\columnwidth]{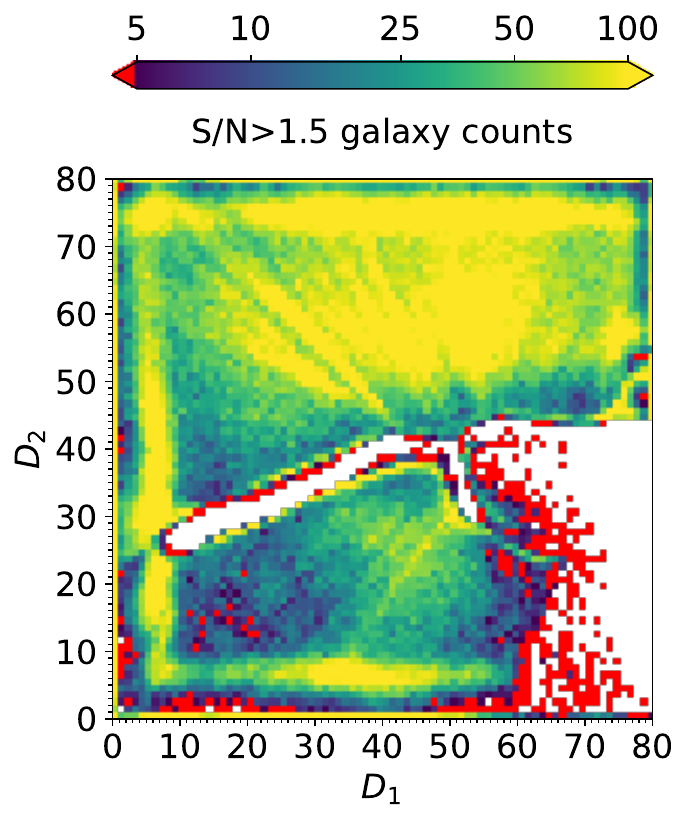}
     \caption{The \textsc{Horizon-AGN} SOM (same as Fig.~3) colour-coded according to the number of galaxies per cell with $S/N>1.5$ in each band (white pixels indicate empty cells). }
     \label{fig:som_occu_snr}
 \end{figure}

 As mentioned in Sect.~4.1, the selection of COSMOS-like galaxies with $S/N>1.5$ may bias the results. 
 First of all, we note that such a selection corresponds in practice to a magnitude cut $i^+<25$  (see Fig.~\ref{fig:som_method_hist}, upper panel) with the removal (because of the required $u$-band detection) of $z>3.5$ galaxies (Fig.~\ref{fig:som_method_hist}, lower panel).
 Even when working in the narrower range $0<z<3.5$, the SOM would still be biased if any  galaxy type were missing from the training sample. To investigate this potential issue we consider the SED classes (i.e., the  cells) defined in the initial SOM (shown in Fig.~3 of the main text) and verify whether they are well-represented after the $S/N$ cut. We find that most of those classes are still included with a fairly good statistics (Fig.~\ref{fig:som_occu_snr}). The occupation is larger than 10 galaxies per cell in most of them. Nearly 10 per cent of the cells, predominantly at $z>3.2$, have less than 5 objects inside; these significant lack of high-$z$ galaxies introduces the $z_\mathrm{SOM}$ bias shown in Fig.~8 of the main text.

\subsection{Calibration sample C1}

  Another step in the SOM method is to label its cells using a sample of \textit{bona fide} galaxies. Sect.~5 presented two possible ways of building such a sample with future telescopes. 
  
  The calibration sample C1 relies on the unparalleled multiplexing that 4MOST and MOONS will offer (placing $10^2$ to $10^3$ fibers simultaneously) in order to observe the several hundreds of \textit{bona fide} galaxies needed to label the SOM cells. 
  Both spectrographs will start operations in 2020$-$2022, at the ESO telescopes VISTA and VLT  respectively. Together they shall enable for H$\alpha$ detection in thousands of galaxies between $z=0$ and $\sim$1.7, which is the highest redshift that MOONS can reach before $H_\alpha$ moves outside its wavelength window (which ends at 1.8\,$\mu$m). 4MOST will observe $H_\alpha$ emission up to $z\sim0.4$, i.e.\ one sixth of the SOM map (see the redshift map in Fig.~7).  
  According to its expected sensitivity, 4MOST should provide $S/N>6$ spectra for galaxies brighter than 20\,mag, in 2h of exposure with ``grey sky'' conditions\footnote{
  Sensitivity per resolution element, converted from the $S/N$ per \emph{\AA} provided by the 4MOST team at \url{https://www.4most.eu/cms/facility/capabilities/}. Sensitivity for emission lines is not provided.}. The spectral resolution  ranges from $\sim$4,000 in the bluest band, up to 7,700. 
  The MOONS Consortium predicts that the sensitivity of their instrument shall reach $S/N=5$ for a line flux of $0.6\times10^{-17}$\,erg\,s$^{-1}$\,cm$^{-1}$, in 1h exposure\footnote{
  Predicted $S/N$ per resolution element including background subtraction, assuming an emission line with FWHM$=$200\,km\,s$^{-1}$ \citep[see \url{https://vltmoons.org/} and ][]{moons_spie18}}.
  
  Concerning the ``observational bias'' introduced by C1, a major concern is the survey strategy of selecting targets among the brightest galaxies in each cell, i.e., objects brighter than  $\langle i^+\rangle^\mathrm{cell}$ to enhance the probability of a $S/N>10$ line detection. 
 The dominant source of error is the aperture correction, which in future surveys may be mitigated by larger slits\footnote{Although this is not the case of either 4MOST or MOONS, which will have fibers with a similar diameter of Subaru FMOS.} or by overlapping with grism or IFU spectroscopy (HST, \emph{Euclid}, JWST) that would allow for an object-by-object correction. 
 We do not attempt to model systematic uncertainties related e.g.\ to the dust prescription (see Eq.~6) or to the conversion factor between nebular and stellar extinction  \citep[see][]{kashino13}.  
 Fig.~11 illustrates the solidity of manifold learning: despite the stochastic SFR$_\mathrm{cal}$ fluctuations due to the random selection of these galaxies, the SOM labels still show a clear evolutionary pattern.

\subsection{Calibration sample C2}
\label{subsec:bias_calib_sample}

 In the case of C2, a  $19\arcmin\times19\arcmin$ patch of UV and FIR
 pseudo-observations is chosen to be the S-W corner of the \textsc{Horizon-AGN} lightcone and supposed to be observed in UV and FIR. 
 Final results are preserved if we change the location of the patch. 
 Compared to the previous calibration, this one has the advantage that more \textit{bona fide} galaxies  are observed in each cell, averaging the intrinsic variance in their star formation activity. Moreover, the redshift range is larger (see \ref{fig:som_method_hist}, bottom panel). 
 
 As the SFR$_\mathrm{cal}$ values come from UV+IR energy balance (Eq.~7 in the main text) precise redshifts for the \textit{bona fide} galaxies should be gathered to compute rest frame  luminosities.   
 This apparently implies an additional follow-up to get spectroscopic redshifts (see discussion in Sect.~5.1). However, following the parallelism between the \textsc{Horizon-AGN} lightcone and COSMOS2015, we can assume that in our field there is already an extended number of them. In the real COSMOS field these $z_\mathrm{spec}$ come from two decades of spectroscopic campaigns, most notably  the ongoing survey for a Complete Calibration of the Color-Redshift Relation \citep[C3R2,][]{masters17}. The C3R2 programme is systematically collecting spectroscopic redshifts in each cell of the COSMOS SOM \citep[already covering $\sim$75 per cent of it,][]{masters19} building on previous surveys as zCOSMOS \citep{lilly07}, VUDS \citep{lefevre15}, Keck/DEIMOS \citep{hasinger18}.

 In a traditional analysis a key concern in the design of C2 would be cosmic variance uncertainty, hereafter referred as ``sample variance''\footnote{Although the term less popular in the literature \citep[see][]{moster11} \textit{sample variance} is more advisable to indicate the uncertainty due to large-scale clustering which changes the observed density of a given type of object. Strictly speaking, \textit{cosmic variance} describes the intrinsic limitation of cosmological experiments, which cannot be reproduced in other observable universes.}. This is due to the fact that the necessary UV and FIR photometry is taken from deep pencil-beam surveys covering only 0.1\,deg$^2$ (10\% of the total area of the \textsc{Horizon-AGN} lightcone). The relatively small area means that specific types of objects may be over-represented and dominate the measured density. Given the quality of our results, sample variance does not seem to dramatically impair our method. Nonetheless, it might bias the SOM in a subtle way. An assessment  of sample variance uncertainties for a given survey design is always recommended, but difficult to pursue because ideally it requires several independent realisations of the survey itself.   
  
  \citet{buchs19} does an extensive analysis of how sample variance would affect mapping functions like the one proposed here, and how to correct for it.  Their analysis relies on multiple realisations of a Gpc-scale numerical simulation (although limited to $z<1.5$). They find that sample variance is a major contributor to the error budget, accounting for $3\!-\!5$ times the shot-noise uncertainty when a calibration sample covers an area of 1\,deg$^2$ (see their figure 4). However, if the parent photometric data set is observed over a large enough area, the variance in the calibration field can be determined and corrected for.  So, in principle sample variance is much less of an issue with the SOM method.  However, we are limited by the box size of \textsc{Horizon-AGN}  and the   $19\arcmin\times19\arcmin$ sub-region we have selected still corresponds to a  small cosmic volume in which some galaxy types may be under-represented and others over-abundant respect to the rest of the sample.  
  For example if a dark matter over-density is present along the line of sight of the pseudo-survey, then a higher fraction of \textit{bona fide} galaxies at that redshift will be ``red and dead'' and perhaps cells of star-forming objects  will not be filled. 
  The bottom panels of Fig.~11 (see main text) shows this concept: even though C2 galaxies span the whole redshift range, a few isolated cells (or small groups of cells) are empty. 
  On the other hand, C1  galaxies are dispersed in a much larger volume and are likely to live in a broader variety of environments (besides the fact they fill a contiguous area of the SOM by construction).  A certain degree of sample variance can be appreciated also in the redshift and magnitude distributions shown in Fig.~\ref{fig:som_method_hist}. 
  We also note that sample variance may be misidentified with other survey selection effects: for instance the SFR map from C2 calibration   differs from C1  not for the smaller area, but mainly because in the latter the ``spectroscopic'' $F_{H\alpha}$ threshold favourites the targeting of more actively star-forming galaxies.

\bsp    
\label{lastpage}
\end{document}